\documentclass[twocolumn%
]{aastex631}

\usepackage{amssymb,amsmath,verbatim,mathtools,needspace,enumitem,etoolbox,graphicx,physics,microtype,afterpage,bigints,gensymb,tabularx,xspace,bm}%
\usepackage{multirow}

\usepackage[normalem]{ulem}

\usepackage{xfrac}
\definecolor{linkcolor}{rgb}{0.0,0.3,0.5}
\definecolor{dodgerblue}{HTML}{1E90FF}
\usepackage[all]{hypcap}
\usepackage[T1]{fontenc}
\usepackage[utf8]{inputenc}
\usepackage{orcidlink}
\usepackage{tensor}
\usepackage{graphicx} %

\makeatletter
\newcommand*{\balancecolsandclearpage}{\close@column@grid \cleardoublepage \twocolumngrid}
\makeatother

\newcommand{\columbia}{\affiliation{Columbia Astrophysics Laboratory, Columbia University, 550 West 120th Street, New York, NY 10027, USA}}

\begin{document}

\title{Pulsar Rockets and  Gaia Neutron Star Binaries}%

\author{Vishal Baibhav$\,$\orcidlink{0000-0002-2536-7752}}
\footnote{NASA Einstein Fellow; \href{mailto:vb2630@columbia.edu}{vb2630@columbia.edu}}
\columbia
\author{Andrei Gruzinov}
\affiliation{CCPP, Physics Department, New York University, 726 Broadway, New York, NY 10003}
\author{Yuri Levin$\,$\orcidlink{0000-0002-6987-1299}}
\affil{Physics Department and Columbia Astrophysics Laboratory, Columbia University, 538 West 120th Street, New York, NY 10027}
\affil{Center for Computational Astrophysics, Flatiron Institute, 162 5th Avenue, 6th floor, New York, NY 10010
}
\affil{Department of Physics and Astronomy, Monash University, Clayton, VIC 3800, Australia}

\pacs{}

\date{\today}

\begin{abstract}

We prove that the spin-aligned electromagnetic recoil force acting on a pulsar vanishes identically in Force-Free Electrodynamics. This contrasts with Hirai et al.'s recent argument that the ``rocket'' effect was important for explaining the eccentricity distribution of wide neutron star binaries found by Gaia. Our detailed analysis confirms that for a broad range of initial conditions and natal kick distributions, the rocket velocities of $v_r \gtrsim 30 \mathrm{km/s}$ are required to account for the observed eccentricities. However, we find that in scenarios where the common envelope phase does not significantly shrink the initial orbit, and the natal kicks are drawn from Paczynski-type  distribution, these eccentricities may arise without the influence of an EM rocket. If the natal kicks are perpendicular to the initial orbits, then explaining the Gaia neutron star eccentricities without invoking rockets additionally requires that the pre-supernova binaries avoid significant circularization, and that the mass loss during the supernova is minimal.

The rocket effect can only be substantial in "weak" pulsars, where pair production near the light cylinder is suppressed and ${\bf E}\cdot{\bf B} \neq 0$ in the outer magnetosphere. We derive a rough estimate for a rocket force 
in a weak pulsar %
and relate it to the pulsar's radiative efficiency;  simulations are needed to obtain numerically reliable expressions.  
If future observations prove that rockets are required to explain the data,  this would imply that Gaia neutron stars are $\gtrsim $ Myr old, were previously rapidly spinning and are weakly magnetized, with a dipole field $\lesssim 10^{10}$ G and a relatively strong quadrupole component.

\end{abstract}

\section{Introduction}

The recent Gaia discovery of 21 wide binaries, each comprising a neutron star and a main-sequence, solar-type companion \citep{2024OJAp....7E..58E}, poses new challenges to standard models of binary evolution. The broad range of eccentricities of these systems is difficult to reconcile with traditional formation scenarios. \citet{Hirai:2024pmx} argued that the electromagnetic (EM) rocket effect, a recoil caused by asymmetric EM radiation and EM-driven outflow, may be required to account for the observed orbital properties. This paper critically revises and extends the observational argument for the rockets in binaries, and radically revises the theory of electromagnetic rockets.

\subsection{Gaia neutron star binaries}
The orbital periods of the binaries are between $190$ and $1000$ days, the masses of the neutron stars are between $1.3 M_\odot$ and $1.9M_\odot$, and the masses of their main sequence companions are between $0.7M_\odot$ and $1.2 M_\odot$. 
No radio pulsations were detected, which is not surprising given the age of the systems: a non-recycled radio pulsar would have spun down below the  ``death line'' in a much shorter time \citep{1993ApJ...402..264C, 2022ARA&A..60..495P}. However, it is reasonable to assume that at the start of their lives, the neutron stars were exhibiting pulsar phenomena, and their rapid rotation was driving powerful electromagnetically-dominated winds.

The eccentricity distribution of the binaries is broad and some binaries have a remarkably low eccentricity. The cumulative distribution in Fig.~\ref{fig:cdf_kick}
shows a roughly uniform distribution between $0$ and $0.8$, with the lowest value of $0.12$ and the highest of $0.795$. As argued by \cite{Hirai:2024pmx} (hereafter H24), one would not naturally expect to find low-eccentricity neutron star binaries at such a wide orbit. The orbital velocities of the binaries are $\sim 40\hbox{km}/\hbox{sec}$, and thus a typical pulsar natal kick of $\sim 400 \hbox{km}/\hbox{sec}$ \citep{2005MNRAS.360..974H, Faucher-Giguere:2005dxp} would destroy the binaries, and even a much smaller kick would strongly perturb them. It is more natural, as H24 argue, for these binaries to originate at much smaller separations than currently observed, in which case the natal kicks are less likely to destroy the binaries; the smaller separations are in fact expected from the common-envelope stage of the binaries' evolution, when the neutron star progenitors expand into giants with their envelopes extending to and beyond the currently observed orbits of their companions. 

\subsection{Observational argument for rockets}
H24 argue that a combination of smaller initial periods, together with natal supernova kicks and Blaauw kicks \citep{1961BAN....15..265B}, would produce strongly eccentric orbits, inconsistent with the currently observed eccentricity distribution. They show, however, that if one takes into account the rocket effect, i.e.~the constant force acting along the pulsar spin  during the spindown, then the rocket force can exert torque on the initially eccentric orbits and substantially circularize them, bringing the eccentricities into a better agreement with the observed broad distribution that includes low values. Building on previous work by \cite{1975ApJ...201..447H} and \cite{Petri:2019xrm}, H24 emphasized that while for an isolated pulsar it is impossible to distinguish observationally the effects of the supernova natal kick and the gradual rocket acceleration, their effects are dramatically different for a pulsar motion in a wide binary. Therefore, the Gaia neutron star binaries could provide the first observational evidence for pulsar rockets.

In Section $2$ of this paper, we use Monte-Carlo simulations of post-supernova orbits, combined with Gaia selection functions for neutron-star binaries, to examine whether the data does require rockets. We find that some very specific, but plausible distributions of the pre-supernova orbital parameters and natal kicks result in the eccentricity distributions which are compatible with the observed ones; these require 1. the pre-supernova binary separations to be of the same order as the currently observed ones, and 2. The natal kick distribution function to have relatively high values at the low-velocity end. Therefore, strictly speaking, currently the data does not require rockets. However, we do find that for a broad class of pre-supernova orbital parameters and the ``allowed'' distribution of natal kicks\footnote{I.e., the distribution compatible with the  proper motions measured for non-recycled pulsars.}, the rocket effect greatly helps to explain the data. These scenarios require rocket velocities of order $30\hbox{km/sec}$, which place very strong constraints on the physics of the Gaia neutron stars.

\subsection{Electromagnetic rocket theory}
 We  find that the rocket strength is much smaller than that obtained in previous investigations. The original proposal of \cite{1975ApJ...201..447H} and subsequent computations \citep{Lai:2000pk, Kojima:2010vt, Petri:2016ncq,Petri:2019xrm, Igoshev:2023ypd} were based on the following consideration: if the dipolar magnetic field is offset from the center, or more generally, if the field has a significant multipolar component, the resulting anisotropic spin-down radiation can create a net force along the spin axis, gradually increasing the NS’s velocity. These calculations showed that for an initially rapidly spinning neutron star, the terminal velocity of the pulsar could be hundreds of km/sec, with particularly high values for nearly-aligned rotators.

However, all of these computations assumed that the pulsar is rotating in vacuum. This is a poor approximation for a pulsar above the death line; their magnetospheres are filled with electron-positron pair plasma that will screen the component of the electric field parallel to the magnetic field in the rotating frame. In those magnetospheres $\vec{E}\cdot\vec{B}\simeq 0$, where $\vec{E}$ and $\vec{B}$ are the electric and magnetic fields respectively. As we will show in Section 3, an exact equality $\vec{E}\cdot\vec{B}=0$, such as in the Force-Free magnetosphere, would lead to the rocket force which is {\bf identically zero}. Our argument is purely analytical, but we note that it is in contradiction with the numerical study of the rocket effect from a rotating Force-Free magnetosphere \citep{2021MNRAS.501.4479P}. We comment on where, in our opinion, the latter study could have gone wrong.

Clearly, the previous computation grossly overestimated the rocket effect, but Force-Free Electrodynamics  is also too idealized since it explicitly does not contain dissipation. We construct a crude estimate of the rocket force acting on a dissipative pulsar,
and obtain an order-of-magnitude scaling of the rocket velocity as a function of the pulsar's initial spin, magnetic field, and age.

In Section 4 we discuss the implications of the above results. We argue that if future observations show that the rockets are required, this would imply that the Gaia neutron stars were borne rapidly rotating, with weak $\lesssim 10^{10}\hbox{G}$ magnetic fields, strongly non-dipolar magnetospheres,
and high radiative efficiencies, similar to currently observed millisecond pulsars.

\section{Gaia Constraints on Natal Kicks and Electromagnetic Rockets}\label{sec:gaia_rockets}

\begin{figure*}[htbp]
    \centering
    \includegraphics[width=\textwidth]{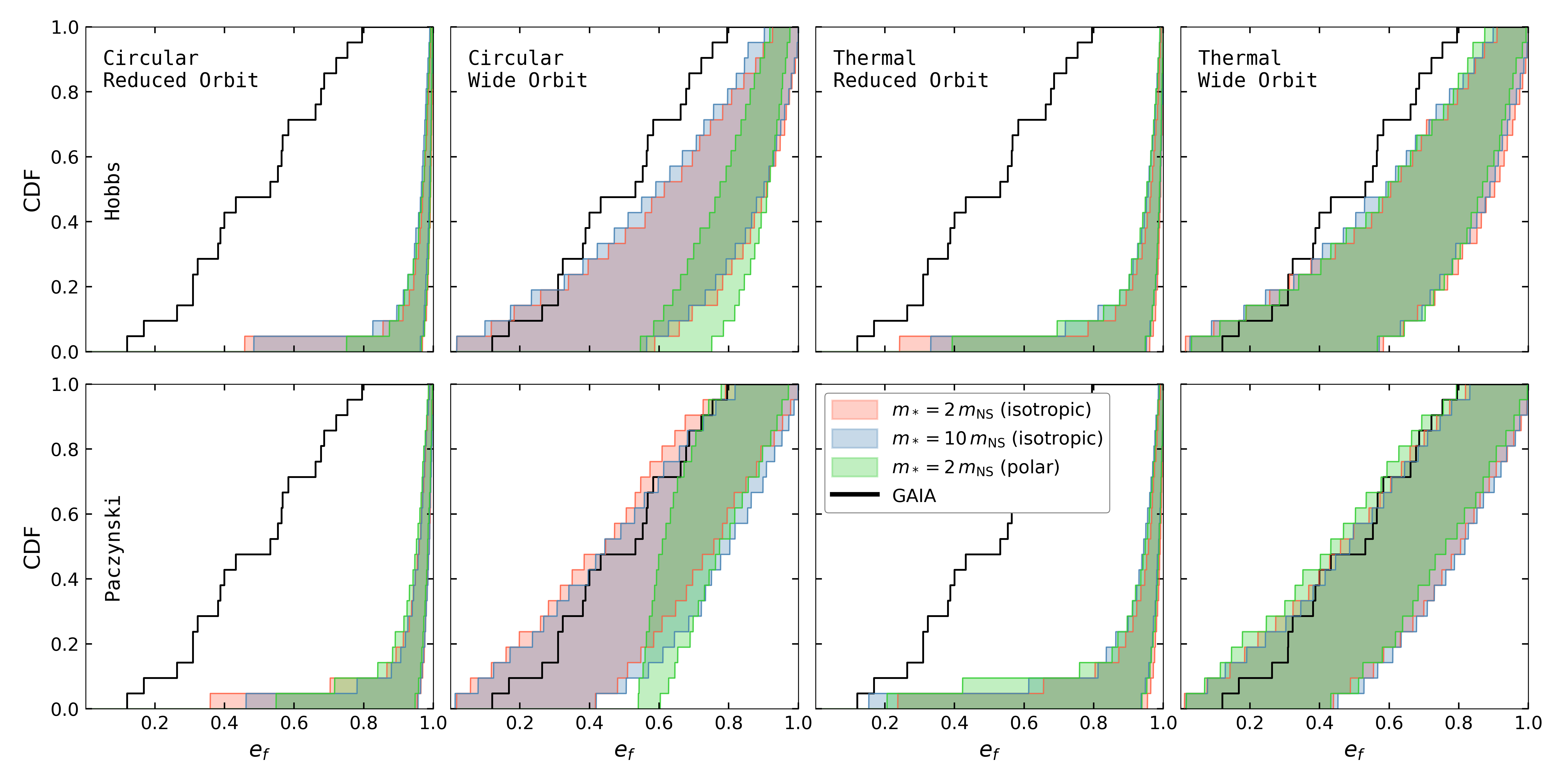}
    \caption{Cumulative eccentricity distributions of neutron star binaries \textit{without rockets} compared with Gaia observations. Each panel corresponds to a combination of initial orbital eccentricity (circular vs. thermal) and separation (reduced vs. wide), organized by kick distribution: Hobbs (top row) and Paczyński (bottom row). The red and blue shaded bands represent the 99\% confidence intervals for model populations with isotropic kicks, corresponding to progenitor masses of $m_* = 2m_{\rm NS}$ (red) and $m_* = 10m_{\rm NS}$ (blue). The green band shows the same for polar kicks with a progenitor mass of $m_* = 2m_{\rm NS}$. The observed Gaia CDF is overplotted in black. Scenarios with wide orbits yield eccentricities most consistent with the observed Gaia distribution, particularly under the Paczyński kicks.}
    \label{fig:cdf_kick}
\end{figure*}

\begin{figure}[htbp]
    \centering
    \includegraphics[width=\columnwidth]{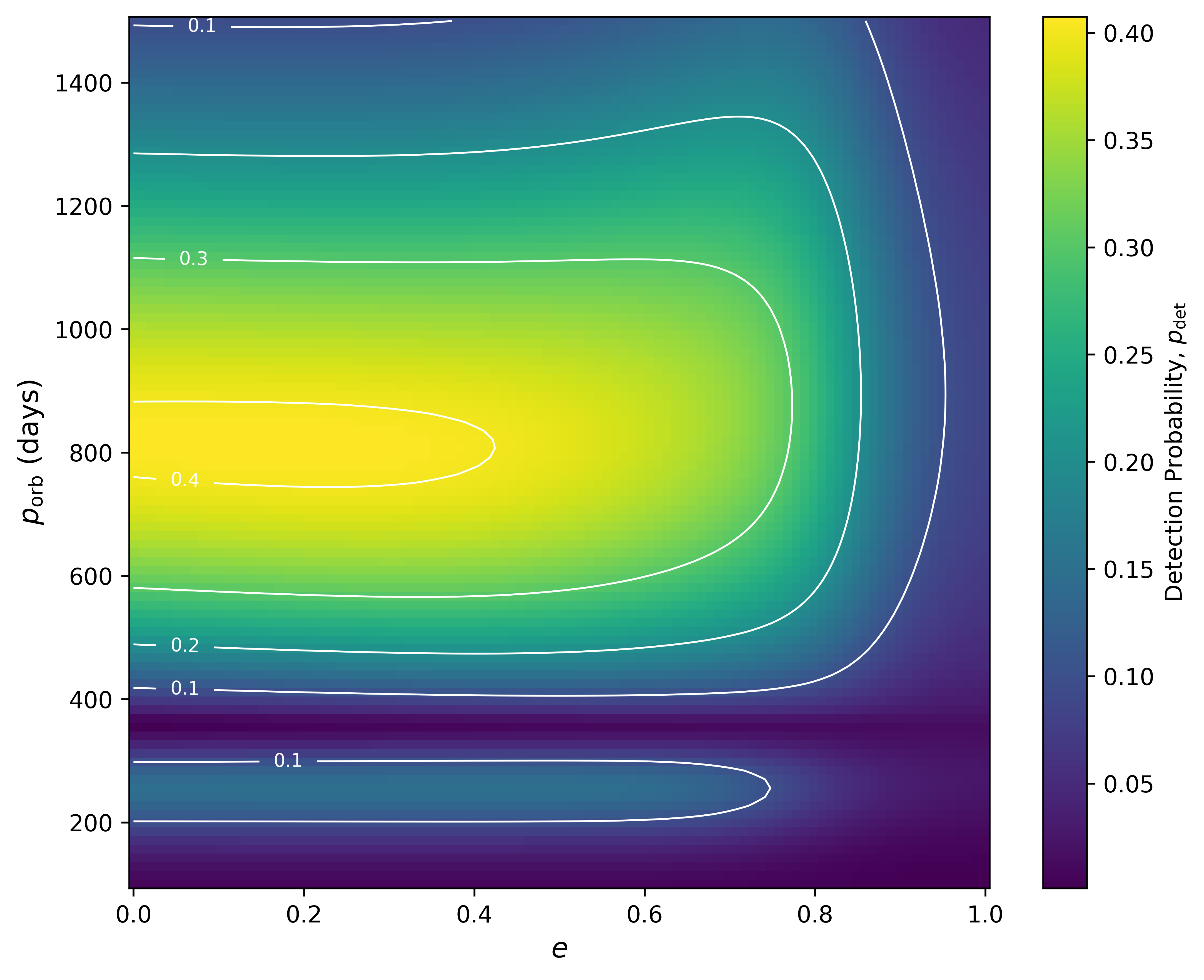}
    \caption{Gaia detection probability $p_{\rm det}(P_{\rm orb}, e)$ as a function of orbital period and eccentricity. }
    \label{fig:p_det_2D}
\end{figure}

The main goal of this section is to explore whether the eccentricity distribution of Gaia neutron star binaries can be explained by a combination of the Blaauw and supernova natal kicks, or if the electromagnetic rocket effect  is required.
To this end, we construct synthetic populations of binaries in which the massive progenitor star undergoes a supernova to form a neutron star. These populations are designed to reflect a range of plausible orbital separations and eccentricities, motivated by current uncertainties in binary stellar evolution.

\textbf{\em Binary initial conditions:} Explaining the Gaia neutron star binaries  is challenging because the observed binaries have orbital separation that are smaller than the predicted radii of neutron star progenitors once they leave the main sequence.   Naively, this would seem to indicate that the progenitor star would have engulfed the solar-type companion prior to supernova. Traditional binary evolution models suggest that such  orbital separations should trigger unstable mass transfer episodes and common envelope (CE) phases that are expected to dramatically shrink the orbits before the supernova. This suggests that the orbits did substantially reduce in size during the CE phase, but subsequently widened due to the natal kicks.  However, this argument might not be on solid ground, because our theoretical understanding of the CE phase is largely based on simplified prescriptions characterizing very complex hydrodynamics. As a case in point, the orbital separations of white dwarf binaries detected by Gaia also appear anomalously large, as these systems are also expected to have undergone significant orbital contraction during a common envelope (CE) phase  \citet{2024OJAp....7E..58E}. However, since white dwarfs do not receive natal kicks, their orbits did not shrink and thus current models of CE phase disagree with observations. 

We therefore must remain agnostic about the pre-supernova separations of Gaia binaries. To explore the range of possibilities, we consider two extreme cases:
\begin{itemize}[label=-,noitemsep, topsep=0pt, leftmargin=*, labelsep=0.5em]
  \item \textit{Wide Orbits:} Initial semi-major axes $a_i$ are drawn from a log-uniform distribution in $[0.1, 10]$ AU, assuming no CE-induced shrinkage. These systems are already wide enough to resemble Gaia-like binaries.
  \item \textit{Reduced Orbits:} $a_i$ are drawn from $[0.01, 0.1]$ AU, representing systems whose orbits were significantly contracted by a CE phase.
\end{itemize}
Similarly, we consider two scenarios for the pre-supernova orbital eccentricity:
\begin{itemize}[label=-,noitemsep, topsep=0pt, leftmargin=*, labelsep=0.5em]
\item \textit{Circular:} $e_i = 0$, assuming circularization during the common envelope.
\item \textit{Thermal:} Eccentricities are drawn from a thermal distribution\footnote{%
We have no particular basis for this assumption, but it is a convenient benchmark to bracket a range of entropy for Gaia binaries. 
}, $p(e_i) = 2e_i$ for $e_i \in [0,1]$.
\end{itemize}

To better understand the influence of natal kick magnitudes on the post-supernova orbital eccentricity distribution, we compare two kick models: the widely used Hobbs distribution and the alternative Paczyński distribution. Both are informed by observed pulsar proper motions but differ significantly in how they represent the distribution of low-velocity kicks.
\begin{itemize}[label=-,noitemsep, topsep=0pt, leftmargin=*, labelsep=0.5em]
    \item \textit{Hobbs Kick Distribution:} The \citet{Hobbs:2005yx} distribution is derived from the proper motions of young isolated pulsars. They argued that these three-dimensional kick velocities are well-described by a Maxwellian distribution with a one-dimensional velocity dispersion of $\sigma_k = 265 \mathrm{km\ s^{-1}}$ (peaking at a three-dimensional speed of approximately $375 \mathrm{km\ s^{-1}}$). The probability density function for the kick speed $v_k$ is given by:
    \begin{equation}
    p(v_k | \sigma_k) = \sqrt{\frac{2}{\pi}} \frac{v_k^2}{\sigma_k^3} \exp\left(-\frac{v_k^2}{2\sigma_k^2}\right)
    \end{equation}
    
    This has become a standard in studies of pulsar kinematics and population synthesis. However, this distribution heavily favors large kicks, with relatively few systems receiving low-magnitude kicks. For example, the probability of a kick speed $v_k < 60\,\mathrm{km/s}$ is $P(v_k < 60\,\mathrm{km/s}) \approx 0.003$ with this distribution.

    Despite its widespread use, the \citet{Hobbs:2005yx} distribution has faced scrutiny. \citet{Verbunt:2017zqi} argue that there are systematic errors in the distance estimates used by \citet{Hobbs:2005yx}. They also show that the transverse velocities of the pulsars analyzed by \citet{Brisken:2002ri} are difficult to explain with this kick distribution. More recent analyses of pulsar kinematics and binary populations (e.g., \citet{Verbunt:2017zqi, Igoshev:2020lif, Disberg:2025xoh}) often favor significantly lower average kick velocities than the Maxwellian found by \citet{Hobbs:2005yx}. Furthermore, \citet{Valli:2025eyd} find that the kick distribution of \citet{Hobbs:2005yx} is strongly disfavored for explaining the properties of their Be X-ray binary sample. \citet{Disberg:2025xoh} point out that this tension could arise because the Jacobian correction for the logarithmic bin sizes used in the original analysis might be missing from \citet{Hobbs:2005yx}. {\citet{Willcox:2021kbg} proposed that stars in an interacting binary produce some neutron stars with small kick velocities, possibly due to Electron-Capture Supernovae.} Because of these issues and the potential biases towards high velocities, we also consider an alternative distribution.

    \item \textit{Paczyński Kick Distribution:} As an alternative that allows for a greater proportion of lower velocity kicks, we consider a distribution based on Paczyński's work, as adopted or modified in later studies (e.g., \citet{1990ApJ...348..485P} and \citet{Faucher-Giguere:2005dxp}). This form proposes a kick velocity distribution that is more heavily weighted toward lower velocities compared to the Hobbs Maxwellian. The natal kick distribution proposed by \citet{1990ApJ...348..485P} is
    \begin{equation}
    p(v_k | \sigma_k) = \frac{4}{\pi \sigma_k \left(1 + \left(\frac{v_k}{\sigma_k}\right)^2\right)^2}
    \end{equation}
    with  $\sigma_k = 560 \mathrm{km\ s^{-1}}$~\citep{Faucher-Giguere:2005dxp}. This distribution allows for a significantly higher fraction of systems with low kick velocities. For example, using this distribution, the likelihood of low-velocity kicks is higher, with $P(v_k < 60\,\mathrm{km/s}) \approx 0.135$~\citep{Faucher-Giguere:2005dxp}. While consistent with the proper motions of isolated pulsars, this distribution is in better agreement with the survival of certain binary systems and the retention of pulsars in globular clusters~\citep{Faucher-Giguere:2005dxp}.
\end{itemize}
{We use the Hobbs and Paczynski distributions to model two extreme possibilities of pulsar kick velocities: the Hobbs distribution represents a scenario with predominantly high kicks, while the Paczynski distribution models a population with lower kicks. In Appendix \ref{sec:transverse}, we  show that the Hobbs model tends to overestimate the observed kick magnitudes, whereas the Paczynski model underestimates them. The evidence therefore suggests the true distribution lies somewhere between the predictions of these two models.}

Finally, we also consider two possibilities for the direction of the natal kick:
\begin{itemize}[label=-,noitemsep, topsep=0pt, leftmargin=*, labelsep=0.5em]
\item \textit{Isotropic Kicks:} The kick direction is randomly oriented, independent of the orbital geometry.
\item \textit{Polar Kicks:} The natal kick is directed along the neutron star’s spin, which we assume to be aligned with the orbital angular momentum before the supernova. This is supported by observations of young pulsars, which often show that their transverse motion in sky is aligned with their spin direction \citep[e.g.,][]{Kramer:2003au, Johnston:2005ka, Wang:2005jg, Ng:2006vh, Johnston:2007gx, Chatterjee:2009ac, Noutsos:2012dt, 2013MNRAS.430.2281N, Mandel:2022sxv}. 
\end{itemize}

The progenitor mass prior to the supernova is uncertain, largely due to poorly constrained supernova physics. To bracket this uncertainty, we fix the progenitor mass to either $m_* = 2m_{\rm NS} $ (representing ``low'' mass loss) or $ m_* = 10m_{\rm NS} $ (representing ``high'' mass loss). These choices allow us to explore the range of possible dynamical outcomes in the binary system. Modest mass loss leads to relatively small changes in orbital separation and eccentricity, whereas large mass losses can induce high eccentricities or even unbind the binary altogether.

 For each binary, the post-SN orbital semi-major axis $a_f$ and eccentricity $e_f$ are calculated from the conservation of energy and angular momentum, incorporating both the mass loss and natal kick. To our sample of generated binaries, we compute and then apply  a Gaia selection function, $ p_{\rm det}(P_{\rm orb}, e_f) $ (described in detail in the Appendix~\ref{sec:pdet}), which captures the fact that binaries with higher eccentricities are much less likely to be classified as binaries by the Gaia team. This selection function was computed by simulating a synthetic population of neutron star binaries distributed throughout the Milky Way and processing them through Gaia’s pipeline. We used the forward-modeling framework developed by \citet{2024OJAp....7E.100E}, which accounts for Gaia’s complex selection effects (including the scanning law geometry, measurement noise, sequential model fitting, and catalog-level quality cuts). In order to carry out our Monte-Carlo simulations,  we  develop an analytical 2-dimensional fitting formula described in the Appendix~\ref{sec:pdet}. 
The detection probability $\hat{p}_{\mathrm{det}}(P_{\mathrm{orb}}, e)$ is shown in Figure~\ref{fig:p_det_2D}. Please note that 1. The binaries with orbital periods near one year exhibit large parallax uncertainties because their orbital motion becomes degenerate with parallactic motion. However, binaries with periods close to one year are more likely to be detected if they have eccentric orbits, since eccentricity helps break the degeneracy between parallax and orbital motion \citep{2024OJAp....7E.100E}.
2. The detection probability decreases as eccentricity increases with low detections for $ e \gtrsim 0.6 $~\citep{2024OJAp....7E.100E}.

\begin{figure}[htbp]
    \centering
    \includegraphics[width=\columnwidth]{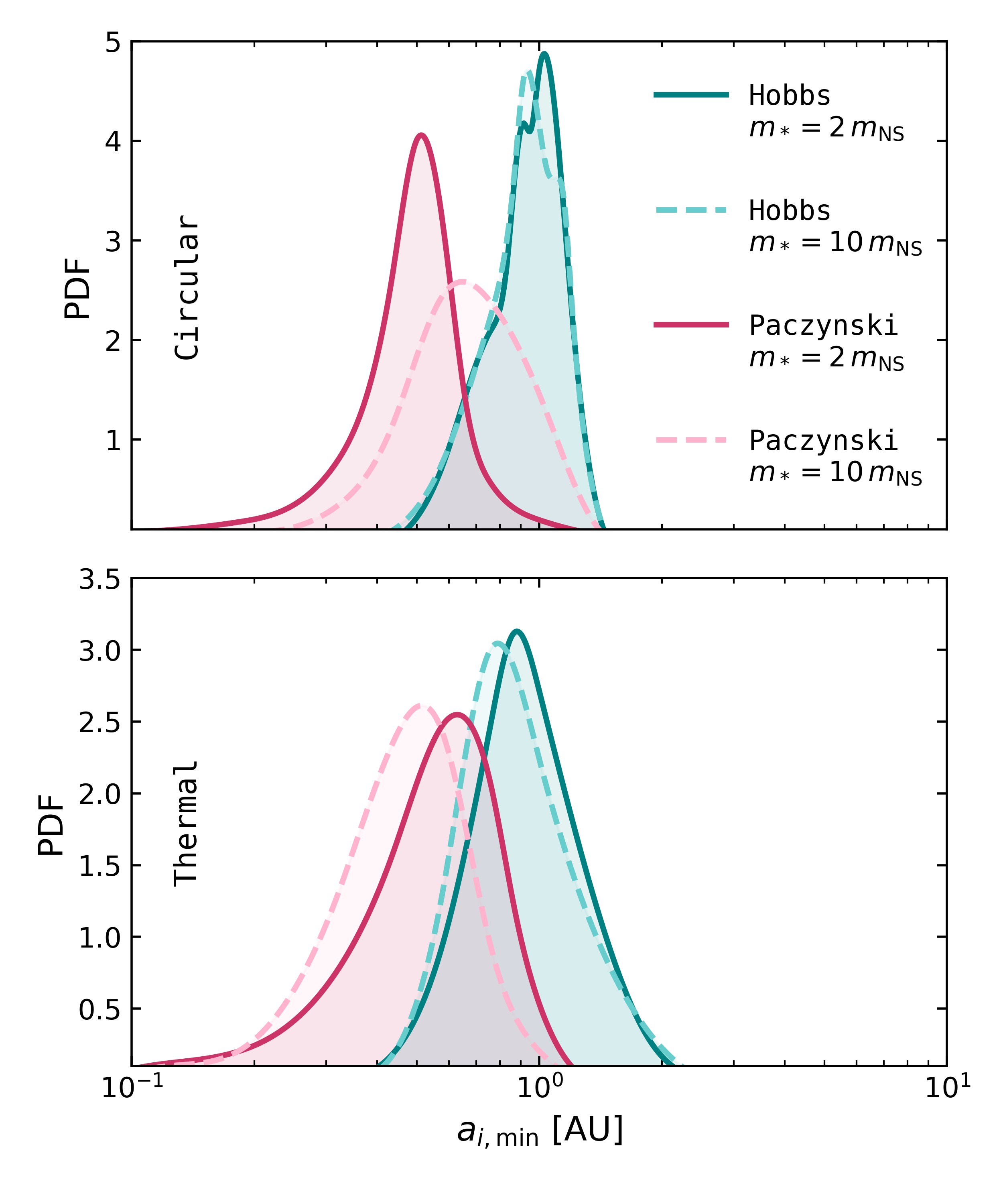}
    \caption{
        Posterior probability density functions of the minimum initial binary separation, $a_{i,\mathrm{min}}$ (AU).
        Each panel corresponds to a different assumed initial eccentricity distribution for the progenitor binary: initially circular orbits (top panel) and initially thermal eccentricity distributions (bottom panel).
        Within each panel, four distinct curves illustrate different combinations of natal kick distributions (Hobbs or Paczynski) and progenitor star masses ($m_*$).
        Solid lines represent a progenitor mass $m_* = 2\,m_{\rm NS}$, while dashed lines correspond to $m_* = 10\,m_{\rm NS}$.
        The colors differentiate the kick distributions: teal shades are used for the Hobbs kick distribution, and pink shades for the Paczynski kick distribution.
        We find that Hobbs kicks generally favor larger minimum initial separations, clustering around approximately $0.5 \text{ to } 1.8\,$AU. In contrast, Paczynski kick distribution tend to favor relatively smaller minimum separations, typically in the range of $0.16 \text{ to } 0.98\,$AU.
    }
    \label{fig:pdf_log_amin}
\end{figure}

\subsection{Can Natal Kicks Alone Explain Gaia’s Observed Eccentricity Distribution?}

 We use our Monte Carlo simulations to construct ``confidence bands'' for the cumulative distribution function (CDF) of the eccentricities. We generate a large number of synthetic populations under the natal kick distribution, and from each population, we repeatedly draw 21 random samples (matching the number of Gaia neutron star binaries). For each sample, we calculate its eccentricity CDF. By aggregating results from millions of such samples, we estimate the pointwise 99\% confidence intervals for the model CDF. These confidence bands allow us to visually and statistically compare the model predictions with the observed CDF to evaluate their consistency.

Fig.~\ref{fig:cdf_kick} compares the observed cumulative eccentricity distribution of Gaia-detected neutron star binaries and synthetic populations generated under various assumptions about natal kicks, progenitor masses, and pre-supernova orbital parameters. Each panel in the figure corresponds to a distinct combination of initial eccentricity (circular or thermal) and orbital separation (reduced or wide), for both the Hobbs (top row) and Paczyński (bottom row) kick distributions.

The red and blue shaded regions show 99\% confidence intervals derived from Monte Carlo populations with progenitor masses of $m_* = 2m_{\rm NS}$  and $m_* = 10m_{\rm NS}$ respectively for isotropic natal kicks, showing the statistical spread in expected eccentricities. The observed Gaia CDF is overplotted in black. 

 The eccentricity distribution is strongly influenced by whether or not CE shrinks the orbit. For the ``reduced orbit" scenario and isotropic Hobbs kick distribution, the median eccentricity is extremely high ($e_f \sim 0.96 - 0.99$), much larger than the observed Gaia median of $\sim 0.5$. This shows the model overpredicts eccentricities when orbits shrink dramatically before supernovae, the point already made in H24. For ``wide orbits", the predicted eccentricities are closer to Gaia observations (median $e_f \sim 0.56 - 0.9$) but still tend toward higher eccentricities than observed.

We repeat the experiment with an isotropic Paczynski-like kick distribution.  For the case with wide initial orbits, the eccentricity distribution shifts towards lower values than those obtained in the scenario with Hobbs' kicks, with medians around $0.4 - 0.8$, much closer to Gaia observations, yielding improved agreement with Gaia. On the other hand, tension still persists for the case with the reduced initial orbits, which produces eccentricities that are still too high to be consistent with the observed values.

The high eccentricities observed in models with reduced orbits can be understood through the following reasoning.
The relationship between the observed eccentricity $e_f$ and the ratio of post- to pre-supernova orbital separations $a_f / a_i$ for initially circular binaries is given by:
\begin{equation}
\frac{1-e_i}{1 + e_f} < \frac{a_f}{a_i} < \frac{1+e_i}{1 - e_f}.
\end{equation}
The distribution of systems tends to cluster near the upper and lower bounds of this relation. Focusing on the upper limit, which is more relevant in scenarios where the orbit expands after the supernova, we can approximate the final eccentricity as $e_f \approx 1 - \frac{a_i}{a_f} (1+e_i)$. This expression explains the different trends in eccentricity distribution for wide and reduced orbits. 
If the orbit was significantly shrunk prior by CE to the explosion, such that $a_i / a_f \ll 1$ (as expected in the reduced orbit scenario following a highly efficient common envelope phase), the resulting eccentricity tends toward unity. 
Conversely, if the orbit remained wide after CE, with $a_i \sim \mathcal{O}(a_f)$ (as expected in the wide orbit scenario), then the resulting eccentricities are more moderate.

\begin{figure*}[htbp]
    \centering
    \includegraphics[width=\textwidth]{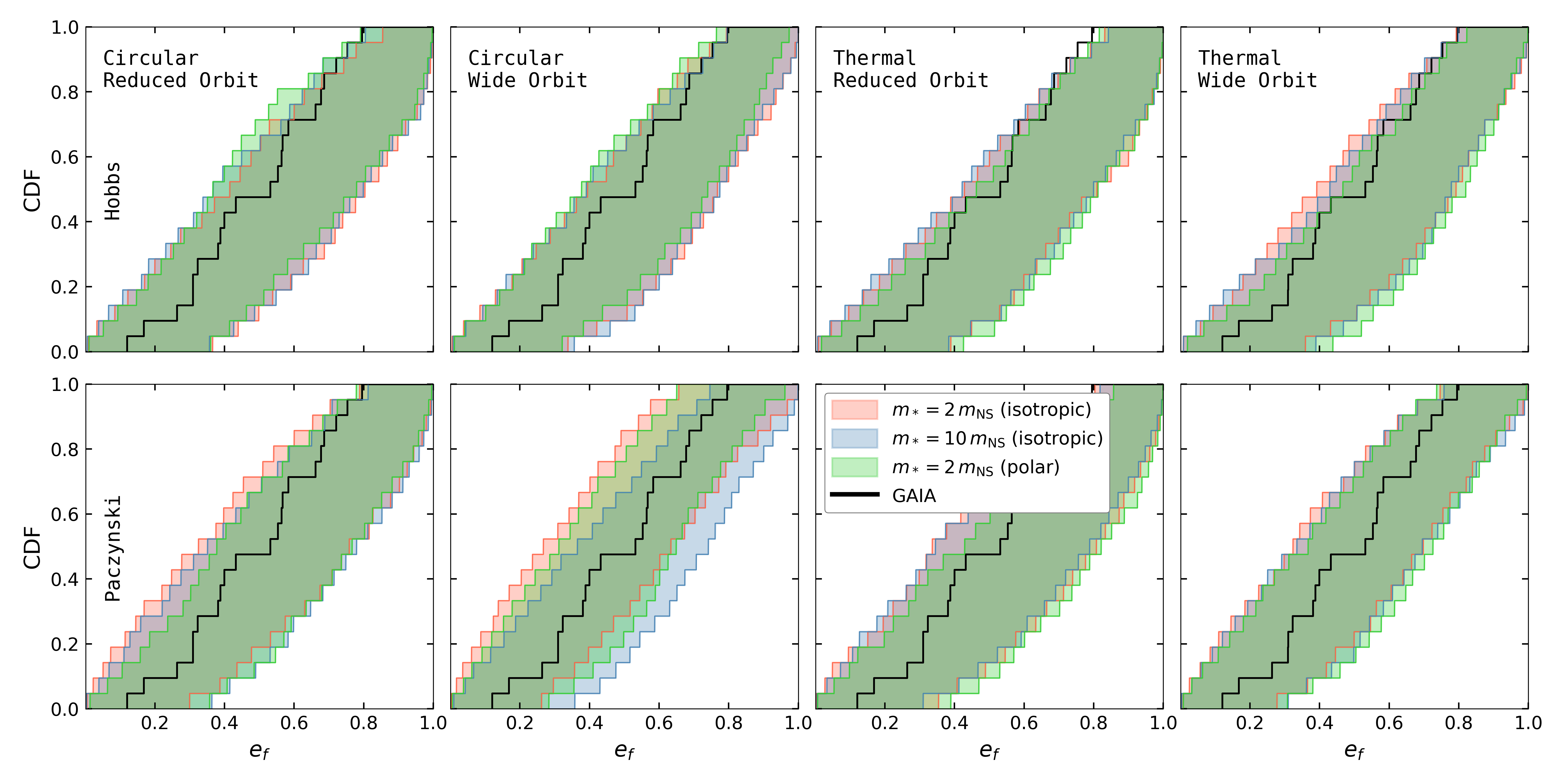}
    \caption{Cumulative eccentricity distributions of neutron star binaries \textit{with rocket} $\sigma_r=50$ km/s assuming isotropic kicks, compared with Gaia observations. Each panel corresponds to a combination of initial orbital eccentricity (circular vs. thermal) and separation (reduced vs. wide), organized by kick distribution: Hobbs (top row) and Paczyński (bottom row). The red and blue shaded bands represent the 99\% confidence intervals for model populations with isotropic kicks, corresponding to progenitor masses of $m_* = 2m_{\rm NS}$ (red) and $m_* = 10m_{\rm NS}$ (blue). The green band shows the same for polar kicks with a progenitor mass of $m_* = 2m_{\rm NS}$. The observed Gaia CDF is overplotted in black. The addition of the rockets significantly improves the agreement with the observed Gaia eccentricity distribution, effectively resolving the tensions seen in the kick-only models as shown in Figure~\ref{fig:cdf_kick}.}
    \label{fig:cdf_rocket}
\end{figure*}

Figure~\ref{fig:cdf_kick} also shows cumulative distribution functions (CDFs)  assuming polar natal kicks and progenitor mass of $m_* = 2\,m_{\rm NS}$ as shown in green. We find polar kicks fail to explain the eccentricity distribution in models with initially reduced orbits as well as in models with initially wide circular orbits. Only systems with initially thermal eccentricities and wide orbits, are able to account for the Gaia eccentricity distribution, when the kicks are assumed to be polar. 
This contrasts with isotropic natal kicks, which could reproduce even circular and wide systems without the need for rockets. This limitation of polar kicks can be understood analytically. For a circular pre-kick orbit and a polar kick, the final eccentricity satisfies
\begin{equation}
e_f > \frac{m_* + m_c}{m_{\rm NS} + m_c} - 1 .
\end{equation}
 Using typical values of $m_{\rm NS} = 1.4\,M_\odot$, $m_c = 1.0\,M_\odot$, and $m_* = 2\,m_{\rm NS}$, above equation implies a minimum eccentricity of $e_f \gtrsim 0.6$, consistent with what is seen in Fig.~\ref{fig:cdf_kick}.

Because the geometry is fixed in the polar case, we can derive exact relations. For circular initial orbits with a polar kick, we find:
\begin{eqnarray}
     a_i &=& a_f (1 - e_f)\\
v_k &=& v_{\rm orb,f} \sqrt{1 - \frac{m_* + m_c}{m_{\rm NS} + m_c}\frac{1}{(1 + e_f)}} .
\end{eqnarray}
 As mass loss decreases, larger kicks are needed to reach the same eccentricity. 
For Gaia NS binaries, the maximum kick typically ranges between $10$–$26$~km/s and $a_i$ smaller than $a_f$ by a factor of $1.1$-$4.9$. In addition,
\begin{equation}
m_* < m_{\rm NS} + e_f (m_c + m_{\rm NS}),
\end{equation}
An important consequence of this lower bound on progenitor mass is that forming low-eccentricity binaries with polar kicks requires extremely small mass loss. For example, consider J1432-1021, which has $e_f = 0.12$ and a neutron star mass of $1.9\,M_\odot$. To achieve this with a polar kick, the progenitor must have been only $m_* = 2.22\,M_\odot$, implying a mass loss of just $0.32\,M_\odot$. This is unusually small for any standard core-collapse scenario.

In summary, natal kicks can reproduce the Gaia eccentricity distribution only if the progenitor binary does not undergo significant orbital shrinkage.
In addition, if the kicks are polar, 
the eccentricities seen in Gaia neutron star binaries are reproduced only with the assumption of extremely small mass loss. If the binaries were circularized before the SN explosion (e.g., because of tidal interaction, the companion with the expanded envelope of the neutron stars' progenitors), then polar natal kicks fail to reproduce the eccentricities of Gaia NS binaries.
In Appendix~\ref{sec:opening_angle} we explore a more general, non-isotropic kick distribution. We find that for initially wide, circular orbits, the kicks biased towards the polar direction, but with a wide opening angle, effectively halfway between strictly polar polar and isotropic, can also successfully reproduce the observed Gaia eccentricities.

\subsubsection{Quantifying the Minimum Binary Separation Required}
\label{sec:bayesian_ai}

Having established that wide binary orbits are a necessary condition for explaining the observed eccentricities in the \textit{Gaia} DR3, we now aim to quantify how wide is ``wide enough.'' Specifically, we constrain the minimum initial orbital separation, $a_{i,\min}$, necessary to reproduce the observed distribution of final eccentricities $e_f$, assuming isotropic natal kicks.  To do this, we constrain the $a_{i,\mathrm{min}}$, using Bayesian inference, with the technical details of our method provided in Appendix~\ref{sec:bayes}.
The posterior probability density functions for the minimum initial binary separation, $a_{i,\mathrm{min}}$, derived from our Bayesian analysis, are shown in Figure~\ref{fig:pdf_log_amin}. The two panels reflect two different assumptions for the initial eccentricity distribution of the progenitor binaries: circular orbits (top panel) and thermal eccentricity distributions (bottom panel). Each panel shows four distinct curves, each representing the posterior PDF for $a_{i,\mathrm{min}}$ under a specific combination of natal kick distribution (Hobbs or Paczynski) and progenitor star mass ($m_* = 2\,m_{\rm NS}$, solid lines, or $m_* = 10\,m_{\rm NS}$, dashed lines). The kick distributions are differentiated by color, with teal shades corresponding to the Hobbs distribution and pink shades to the Paczynski distribution. These posteriors allow us to quantify the most probable ranges for $a_{i,\mathrm{min}}$ consistent with the observed Gaia data under each set of model assumptions.

 The posterior for $a_{i,\mathrm{min}}$ generally peaks around $1\,$AU, with 90\% credible intervals approximately spanning $[0.5, 3]\,$AU. Below $a_{i,\min} \sim 0.5\,\mathrm{AU}$, the predicted final eccentricities become too large compared to observations. On the other hand, increasing $a_{i,\min}$ beyond $\sim 3\,\mathrm{AU}$ dramatically reduces the fraction of binaries surviving the supernova kicks, since wider binaries are more easily disrupted. This trade-off implies a constrained initial orbital separation regime where binaries remain sufficiently wide to reproduce observed eccentricities but not so wide as to be destroyed by kicks.

To conclude, if processes such as common envelope evolution or other binary interactions significantly alter the orbital separations, they must not reduce the separations below the constrained minimum of $\sim 0.5\,\mathrm{AU}$. Bringing binaries closer than this threshold would lead to final eccentricities inconsistent with the observed {Gaia} DR3 distribution, or cause excessive binary disruption.

\subsection{Do Rockets provide better explanation?}

We address this question by repeating the Monte-Carlo simulations, but adding the effect of the electromagnetic rocket aligned with the neutron star spin (assumed to be, in turm, aligned with the original binary's orbital plane). This changes orientation and eccentricity of the binary, which we model using the convenient equations derived in H24. The rocket effect does not change the binaries' semimajor axes.

We draw the rocket velocity $v_r$ from a truncated normal 
distribution with scale $\sigma_r$ and $v_r > 0$: $v_r \sim \mathcal{N}^+(0, \sigma_r^2)$. We assume the commonly held notion that the neutron star's spin is initially aligned with the binary’s orbital angular momentum prior to the supernova [but see \citep{Baibhav:2024rkn} for possible caveats].  When the supernova occurs, the natal kick alters the direction of the orbital angular momentum. As a result, the post-supernova orbital plane becomes tilted relative to the neutron star’s spin axis, with the degree of misalignment set by the kick’s magnitude and geometry.

We show the impact of rocket effect on the final eccentricity distribution in Figure~\ref{fig:cdf_rocket} with $\sigma_r=50$ km/s. 
In all cases, the observed eccentricities lie well within the 99\% confidence bands, showing that the rocket effect broadens the range of possible eccentricities and better matches the observed distribution, especially for scenarios where kicks alone struggle.
 Paczyński  kicks along with the rockets (Fig.~\ref{fig:cdf_rocket}, bottom) yield the best match. Lower kicks avoid excessive $e_f$, and the rocket gently adjusts orbits to fit Gaia’s distribution precisely\footnote{{In Fig. \ref{fig:cdf_kick} and \ref{fig:cdf_rocket},  we compare only the eccentricity from our models with the observed data. Another important observable that is strongly correlated with eccentricity is the final semimajor axis, $a_f$. We find that, after applying Gaia selection effects, the values of $a_f$ predicted by all our models remain consistently in agreement with the observations.
}
}. This is consistent with H24, who argued that rockets are essential in explaining eccentricities if common envelopes shrink the binary orbits prior to supernovae.

\begin{figure}[htbp]
    \centering
    \includegraphics[width=\columnwidth]{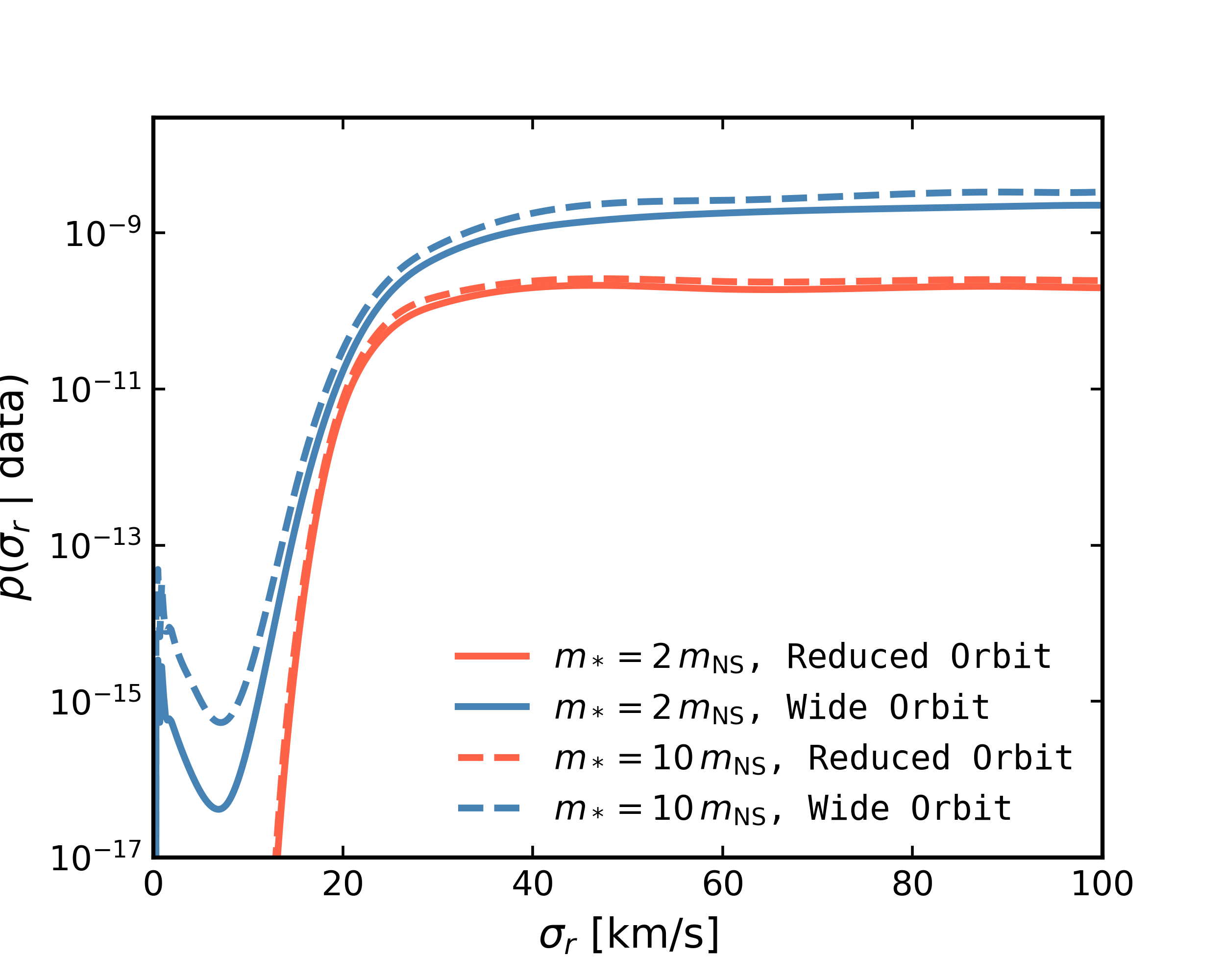}
    \caption{ Likelihood for the rocket magnitude $\sigma_r$ inferred from fitting observed orbital parameters $\{a_{f\rm obs}, e_{f\rm obs}\}$ using the Hobbs kick distribution and circular initial eccentricity. Solid lines show results for progenitor mass $m_* = 2\, m_{\rm NS}$, while dashed lines correspond to $m_* = 10\, m_{\rm NS}$. Colors indicate initial orbital separation regimes: red for reduced orbits and  blue for wide orbits. }
    \label{fig:p_sigma}
\end{figure}

\subsubsection{Inferring the Rocket Kick Magnitude}
\label{sec:bayesian_sigmar}

In this section, we use Bayesian inference to compute the posterior probability density  $p(\sigma_r \mid \{e_f^{(j)}, a_f^{(j)}\})$, which quantifies how likely each value of $\sigma_r$ is, given the data set. {The Bayesian inference methodology, similar to the one employed in Sec.~\ref{sec:bayesian_ai}, is detailed in Appendix~\ref{sec:bayes}.}. Our prior on the rocket velocity dispersion $\sigma_r$ is taken to be proportional to the fraction of detectable binaries as a function of $\sigma_r$ (therefore, the prior is not flat since the detectability depends on the eccentricities, and the eccentricities depend on $\sigma_r$).

Figure~\ref{fig:p_sigma} shows the resulting posterior probability $p(\sigma_r | \mathrm{data})$ under the assumptions of Hobbs natal kick distribution, initially circular pre-supernova orbits for "reduced" and "wide" orbits.
From the figure, it is clear that all models benefit enormously from the rockets with $\sigma_r \ge 30\,\mathrm{km/s}$, and that for the models with reduced initial orbits, zero rocket velocity is entirely excluded. This is because, for these tighter configurations, natal kicks alone produce eccentricities far too high to match the Gaia data. The introduction of a rocket phase moderates these extreme eccentricities, causing the posterior probability to rise sharply until the effect saturates around $\sigma_r \approx 30\,\mathrm{km/s}$, beyond which further increases do not significantly improve the model fit. Bayes factors comparing models with and without the rocket effect decisively favor its inclusion in all scenarios. This support is most pronounced for reduced orbits, but even for wide orbits, the evidence is strong. As shown in Figure~\ref{fig:p_sigma}, this is evidenced by the several orders-of-magnitude difference in posterior probability between the no-rocket ($\sigma_r = 0$) and rocket ($\sigma_r \gtrsim 50\,\mathrm{km/s}$) cases. {Beyond constraining population-level parameters, this Bayesian methodology can also be applied to individual systems. In Appendix~\ref{sec:individual_rockets}, we use this framework to infer the specific rocket kick magnitudes received by each neutron star detected by Gaia.}

\section{Computation of the rocket force from pulsar electrodynamics}
\subsection{General Remarks}
Asymmetric magnetic field on the surface of a {\it rotating} neutron star must give rise to a net force acting on the star. For a star rotating in vacuum, this force can be straightforwardly computed to quadrupole order, as is shown in \S $72$, Problem $2$ of \cite{1975ctf..book.....L}. Astrophysicists use the vacuum formula to interpret various observations.

The vacuum formula is not expected to apply to real pulsars. The pulsar rocket force is large enough to be relevant for observations only for young fast pulsars; young pulsars possess magnetospheres filled with pair plasma, and they are certainly not in vacuum \citep{1969ApJ...157..869G, 1975ApJ...196...51R, 2022ARA&A..60..495P}. But in the past, one assumed that the vacuum expression gives the right {\it estimate} of the pulsar force.

Indeed, consider a closely related problem, the pulsar spindown power. In vacuum, the pulsar power is 
\begin{equation}
L_{\rm vac}=\frac{2}{3}\frac{\mu^2 \Omega^4}{c^3}\sin ^2\theta,
\label{lvacuum}
\end{equation}
where $\mu$ is the dipole moment, $\Omega$ is the angular velocity, $\theta$ is the spin-dipole angle, $c$ is the speed of light. Force-free electrodynamics (FFE) is a much better description of the strong pulsar magnetosphere than vacuum \citep{1999ApJ...511..351C, 2005PhRvL..94b1101G}. In FFE, the expression or the spindown luminosity is given by

\begin{equation}
L_{\rm FFE}\approx \frac{\mu^2 \Omega^4}{c^3}(1+\sin ^2\theta)
\label{spitkovsky}
\end{equation}
 \citep{2006ApJ...648L..51S}. For %
$\cos \theta =1/2$, 
the FFE formula gives a spindown power of 3.5 times larger than that of a dipole in vacuum, but the vacuum formula also works to an order of magnitude.

Previous researchers assumed that the pulsar force, just as the pulsar spindown power, could be computed in vacuum. By this logic, the FFE force is assumed to be different from the vacuum value by a factor of order unity or, at most, a few. This assumption is false: we show in the next subsection that the pulsar force vanishes identically in FFE.

Real pulsars are not exactly in an FFE state, or they would be invisible since FFE does not explicitly include any dissipation. For a non-FFE, non-ideal, dissipative pulsar, there is a finite pulsar force; we give a crude estimate %
of the force formula in Section 3.3. 

\subsection{Zero rocket force in FFE}
The disappearance of the pulsar force in FFE follows from the theoretical description of an FFE pulsar given in \cite{2006ApJ...647L.119G}. We repeat the essential part of the analysis here. In Section 3.3, we will modify some steps and obtain a non-zero force for a dissipative magnetosphere. (In this section for convenience, we use rationalized units and $c=1$.)

For a magnetosphere created by a uniformly rotating star, one assumes stationarity in a rotating frame: 
\begin{equation}\label{E:stat1}
\partial_t{\bf B}=({\bf B}\cdot\nabla){\bf V}-({\bf V}\cdot\nabla){\bf B},
\end{equation}
where
\begin{equation}
{\bf V}\equiv \boldsymbol\Omega \times {\bf r}.
\end{equation}
This equation represents a statement that magnetic field lines are advected by the velocity field $\bf V$, i.e. they are uniformly rotating with the angular velocity $\bf \Omega$. 
Since ${\bf B}$ is solenoidal, we can rewrite Eq.(\ref{E:stat1}) as 
\begin{equation}\label{E:stat2}
\partial_t{\bf B}=\nabla\times({\bf V}\times {\bf B}).
\end{equation}
Now, using Faraday law,
\begin{equation}
\partial_t {\bf B}=-\nabla\times{\bf E},
\end{equation}
we get 
\begin{equation}
\nabla\times({\bf E}+{\bf V}\times {\bf B})=0,
\end{equation}
and therefore
\begin{equation}\label{E:chi}
{\bf E}+{\bf V}\times {\bf B}=\nabla \chi
\end{equation}
for some scalar field $\chi$.
In FFE, %
\begin{equation}
{\bf E}\cdot{\bf B}=0.
\end{equation}
As we will shall see in the following, this this equation implies that the rocket force vanishes.
First we observe that Eq.(\ref{E:chi}) gives
\begin{equation}\label{E:chiffe}
{\bf B}\cdot\nabla \chi=0.
\end{equation}

Second, we note that at the surface of the star, the tangential component of the electric field in a rotating frame vanishes, ${\bf r}\times ({\bf E}+{\bf V}\times {\bf B})=0$, $\chi$ is constant on the surface, and with no loss of generality,
\begin{equation}\label{E:chibc}
\chi|_{r=r_s}=0,
\end{equation}
where $r_s$ is the neutron star radius.

Now assume that all magnetic field lines are anchored to the star at least at one point: a field line can go from the star to the star or from the star to infinity. This seems likely on physical grounds: (i) a pocket of unanchored field should fly away to infinity, (ii) only under this assumption the electric field of the stationary magnetosphere minimizes energy in the rotating frame \citep{2006ApJ...647L.119G}. Numerical simulations, for example, in \cite{2006ApJ...648L..51S}, also show this intuitively plausible magnetic topology. Then Eqs.~(\ref{E:chiffe}, \ref{E:chibc}) give $\chi=0$, and 
\begin{equation}
{\bf E}+{\bf V}\times {\bf B}=0,
\end{equation}
everywhere outside the star.

At the light cylinder (at $\rho=1/\Omega$, in cylindrical coordinates $(\rho, \phi, z)$, with the z-axis aligned with $\boldsymbol\Omega$) we have
\begin{equation}
{\bf E}=-{\bf V}\times {\bf B}=-(0,1,0)\times {\bf B}=(-B_z,0,B_\rho),
\end{equation}
the $z\rho$ component of the Maxwell stress vanishes
\begin{equation}
\sigma_{z\rho}=E_zE_\rho+B_zB_\rho=0%
\end{equation}
There is no z-momentum flux through the light cylinder, and the time-averaged force on the pulsar vanishes identically. A force perpendicular to the rotation axis may exist, but it rotates with the star and its time-average is zero. 

We note that this result contradicts the conclusions in \cite{2021MNRAS.501.4479P}, who performed numerical experiments on time-dependent Force-Free fields sourced by a rotating off-centered dipole, using Maxwell's equations with Ohm's law derived in \cite{1999astro.ph..2288G}. Petri found a rocket force comparable to (in fact, stronger than) that in the vacuum case. Given that our proof is quite rigorous, we are confident that these conclusions are in error. It is hard for us to say exactly where the error is, but we can see one potential weak point. Petri computed the $z$ component of the Poynting vector integrated over a closed sphere [equation (12) of their paper], as a proxy for the $z$ component of the force acting on the pulsar. However, strictly speaking, this procedure is valid only at  asymptotic infinity, and not in the near-zone of the pulsar (the sphere's radius was that of the light cylinder). In the latter case, one needs to work directly with the electromagnetic stress tensor.

\subsection{The rocket force on a %
pulsar with nonzero ${\bf E}\cdot {\bf B}$}
The magnetosphere of a uniformly rotating star is stationary (at least statistically) in a rotating frame, and Eqs.~(\ref{E:chi}, \ref{E:chibc}) still apply. But Eq.~(\ref{E:chiffe}) now reads
\begin{equation}\label{E:chid}
{\bf B}\cdot\nabla \chi={\bf E}\cdot{\bf B}.
\end{equation}
In a non-FFE magnetosphere, in vacuum or in a magnetosphere of a dissipative pulsar, ${\bf E}\cdot{\bf B}\neq 0$ and there will be a net force on a pulsar. We will describe the exact mathematical structure of the force formula in vacuum, (the result that is already known from previous work, but which we recast in a certain convenient form) and give a crude estimate of the force in a dissipative magnetosphere in a particular approximation.

\paragraph{\large Vacuum.}
The force is along the rotation axis,
\begin{equation}
{\bf F}=\alpha \boldsymbol\Omega,
\label{F}
\end{equation}
where $\alpha$ is a pseudoscalar.
For any $r\gg c/{\Omega}$, that is far in the wave zone, $F=-r^2\langle B^2(\hat{z}\cdot\hat{r})\rangle$, where the average is over the sphere $r=$const. To order of magnitude, one can use this expression at the the light cylinder, at $r\sim R\equiv {c}/{\Omega}$:
\begin{equation}
F\sim R^2\langle B^2(\hat{z}\cdot\hat{r}) \rangle.
\end{equation}
To leading order, for a not too fast pulsar, with $R\gg r_s$, the field at the light cylinder is dominated by the dipole component $B_d\sim \mu/{R^3}$, but for a pure dipole the average $\langle B^2(\hat{z}\cdot\hat{r})\rangle $ vanishes. We must include the subleading quadrupole, $B_q\sim q/{R^4}$, where $q_{ab}$ is the magnetic quadrupole pseudo-tensor. One obtains
\begin{equation}
F\sim R^2B_dB_q\sim \frac{1}{c^5}\Omega^5\mu q,
\end{equation}
or
\begin{equation}\label{E:est}
\alpha \sim \frac{1}{c^5}\Omega^4\mu q.
\end{equation}
The mathematical dependence of $\alpha$ on the pseudo-vectors vectors ${\bf \Omega}$, ${\bf \mu}$ and the pseudo-tensor ${\bf q}$  can be deduced on symmetry grounds, because there is only one way to form a pseudo-scalar bilinear in $\mu_a$ and $q_{ab}$. The exact answer is
\begin{equation}
\alpha =\frac{8}{45}\frac{1}{c^5}\Omega^2e_{abc}\Omega_a\mu_bq_{cd}\Omega_d.
\label{structure}
\end{equation}
Here, the numerical prefactor is computed straightforwardly using the result obtained in problem $2$ of \S $72$ of \cite{1975ctf..book.....L}\footnote{LL derive the general expression for the force in terms of the electric dipole $d_a$, the electric quadrupole $D_{bc}$, and the magnetic dipole $m_d$. One readily obtains an expression for the force acting on a rotating electric dipole by noting than in this case $\dot{m}_a=-{1\over 6} \dot{D}_{ab}\Omega_b$ (the Einstein summation convention is assumed), substituting this into the LL's expression for the force, and dropping their erroneous prefactor $(4\pi)^{-1}$. The solution for the rotating magnetic dipole is then trivially obtained from the electric--magnetic duality, i.e. by substituting $\mu_a$ instead of $d_a$ and $q_{bc}$ instead of $d_{bc}$.}.   

\cite{Lai:2000pk} consider a dipole ${\bf \mu}$ displaced by ${\bf s}$ from the center of the star. In this case, 
\begin{equation}
  q_{cd}=3s_c \mu_d+3\mu_c s_d- 2({\bf \mu}\cdot {\bf s})\delta_{cd}.  
\end{equation}
Substituting this into Eqs.~(\ref{structure}) and (\ref{F}), we obtain the expression quoted in \cite{Lai:2000pk}:
\begin{equation}
F={8\Omega^5\over 15 c^5}\mu_z\mu_\phi s_\rho.
\label{fvacuum}
\end{equation}
For a neutron star initially spinning with frequency $\nu_0$, the final rocket velocity is given by
\begin{equation}
v_r={I\over M}\int_0^{2\pi\nu_0}{\Omega F\over  L} d\Omega,
\label{vrocket}
\end{equation}
where the force $F$ and the spindown luminocity $L$ are functions of $\Omega$, and $I$ amd $M$ are the moment of inertia and the mass of the neutron star, respectively. Taking $I\simeq{2/5}MR_0^2$, where $R_0$ is the neutron-star radius, using the expressions for the vacuum force Eq.~(\ref{fvacuum}) and the vacuum spindown luminosity Eq.~(\ref{lvacuum}), we obtain
\begin{equation}
v_r\simeq 180 R_{0, 12}^2 s_{\rho, 12} {\mu_z \mu_\phi\over \mu_\phi^2+\mu_\rho^2}~\nu_{0,716}^3~\hbox{km/s}.
\label{vvacuum}
\end{equation}
Here we normalize $R_0$ and $s_\rho$ to $12\hbox{km}$, and $\nu_0$ to $716\hbox{Hz}$, the frequency of the spin record-holding pulsar, PSR J1748--2446ad \citep{2006Sci...311.1901H}.

However, this equation grossly overestimates the rocket velocity for real pulsars. Firstly, the rocket force $F$ is greatly reduced in nearly Force-Free magnetospheres. Secondly, $L_{\rm FFE}/L_{\rm vac}\approx 1.5(1+1/\sin^2\theta)>3$ and thus the real spin-down luminosity in Eq.~(\ref{vrocket}) is substantially greater than the vacuum estimate, especially for nearly aligned rotators\footnote{\cite{Lai:2000pk} obtain very high terminal velocity for nearly aligned rotators, because the vacuum spin-down luminosity is very small for aligned rotators and is dominated by quadrupole radiation. However, it has been known since \cite{1969ApJ...157..869G} that the spin-down power of aligned rotators is of the same order of magnitude as that of non-aligned rotators because of the contribution of the hydromagnetic wind powered by the rotational energy of the pulsar.}. We give a rough, but more realistic estimate in the following.

\paragraph{\large Dissipative magnetosphere.} The estimate Eq.~(\ref{E:est}) works in vacuum because the ``non-  ${\bf v}\times {\bf B}$'' part of the electric field, $\nabla \chi$, is of the order $B$ near the light cylinder, and there is no precise electric-magnetic cancelation in the Maxwell stress $\sigma _{z\rho}$. But in a nearly FFE, weakly non-ideal dissipative pulsar magnetosphere, the electric-magnetic near-cancellation in $\sigma_{z\rho}$ is important and must be carefully considered.

From Eq.(\ref{E:chid}),
\begin{equation}\label{E:chid1}
{\bf B}\cdot\nabla \chi=E_0B_0,
\end{equation}
where $E_0$, $B_0$ are the proper electric field scalar and magnetic field pseudoscalar. Generally $E_0$ has to be evaluated numerically, and several models exist in the literature. \cite{2013arXiv1303.4094G, 2023arXiv231013638G} solved the magnetosphere using ``Arisotelian electrodynamics '', in which the charge velocity is determined by the local electromagnetic field and the curvature of the field line. Several groups, e.g.
\citet{2014ApJ...795L..22C, 2015ApJ...801L..19P, 2020ApJ...889...69C, 2022ApJ...939...42H, 2023ApJ...958L...9B, 2023ApJ...943..105H, 2024A&A...690A.229C} have performed Particle-In-Cell numerical experiments that take into account pair production in the magnetosphere using phenomenological prescriptions.  

Although the two approaches have not converged, some of the features of the magnetospheres they obtain are common: 1. Much of the dissipation takes place near the light cylinder 2. With an abundant pair production near the light cylinder (the so-called ``strong'' pulsar), the  non-zero ${\bf E}\cdot {\bf B}$ regions have relatively small volume (restricted to gaps and thin current sheets), so the dynamics is expected to be well-approximated by FFE, 3. The pulsars where the pair production near the light-cylinder is not efficient [the so-called ``weak pulsars'' by the terminology introduced in \cite{2015arXiv150305158G}], the dissipation region, while still loosely located near the light cylinder, is extended and occupies the volume of $\sim R^3$. Our estimate of the rocket effect applies to the latter case.

With non-zero $E_0$, a radiation-overdamped (Aristotelian) electron or positron emits gamma rays at a power $ceE_0$. The resulting gamma-ray luminosity of the pulsar, estimated at the light cylinder, is 
\begin{equation}\label{E:dis}
L_\gamma \sim nR^3ecE_0,
\end{equation}
where 
\begin{equation}
n\sim \frac{B}{eR}
\end{equation}
is the Goldreich-Julian density, giving
\begin{equation}
L_\gamma \sim cR^2E_0B.
\end{equation}
The spindown power is
\begin{equation}
L\sim cR^2B^2,
\end{equation}
the radiative efficiency is
\begin{equation}
\epsilon \equiv \frac{L_\gamma}{L},
\end{equation}
and then the non-${\bf v}\times {\bf B}$  electric field at the light cylinder can be estimated as
\begin{equation}
|\nabla \chi|\sim E_0\sim \epsilon B.
\end{equation}

Now we can write down the force formula for a dissipative pulsar:
\begin{equation}
{ F}\sim E_0 B R^2 f ,
\end{equation}
where
\begin{equation}\label{E:alf}
f\sim {q\over R\mu}
\end{equation}
is a factor reflecting the degree of asymmetry at the light cylinder. Putting all of the above together, we obtain an estimate for the rocket force
\begin{equation}
F\sim \frac{1}{c^5}\epsilon \Omega^5\mu q\sim f L_\gamma/c.
\label{force}
\end{equation}

At $\epsilon \ll 1$, a non-trivial geometry of dissipation (almost singular currents sheets) will probably invalidate the estimate Eq.(\ref{E:dis}), and the rocket effect will be strongly suppressed. We also note that we are not able to replicate the beautiful mathematical structure of Eq.~(\ref{structure}) for dissipative pulsars, since the rocket force is not expected to be strictly linear with respect to the pseudo-vector ${\bf \mu}$.

So far, our estimate only counts the momentum flux due to Maxwell stress of the large-scale EM field. There will be also a force due to momentum flux in gamma rays. %
However, to order of magnitude, it is given by the same expression as in Eq.~(\ref{force}), and therefore does not change qualitatively 
our estimate for the rocket force.

We are now ready to give a very rough estimate. In Eq.~(\ref{vrocket}), we multiply the vacuum force by the radiative efficiency $\epsilon$, and multiply the vacuum spin-down luminosity by a factor of $3$ to represent the greater spin-down torque from the hydromagnetic wind. We take the $\mu$-dependent factor from Eq.~(\ref{vvacuum}) to be $1/2$, assuming that all components of $\mu$ are the same. We get
\begin{equation}
v_r\sim 30 \epsilon~\epsilon_q~\nu_{0, 716}^3~\hbox{km/s}.
\label{maximumrocket}
\end{equation}
We have inserted a factor $\epsilon_q$, the ratio of the quadrupole to the dipole field at the surface of the star. While a typical efficiency for weak pulsars is $\epsilon\sim 0.2$ [see teble 14 of \cite{2023ApJ...958..191S}, and \cite{2013arXiv1310.5382G}], they may also have the ``quadrupolar'' $\epsilon_q$ of order several, as observed by NICER in a millisecond pulsar PSR J0030+0451 \citep{Miller:2019cac, Riley:2019yda}. It is not unreasonable to assume that $\epsilon \epsilon_q$ could be of order $1$.

Clearly, to produce effective rockets, the neutron stars had to be very rapidly rotating at their birth. Is it feasible for such rapidly rotating neutron stars to be {\it weak} pulsars? 

Significant pair production near the light cylinder is avoided if the temperature of the star is low enough. According to \cite{2023arXiv231013638G}, the necessary condition for ``weakness'' is
\begin{equation}\label{E:wpul}
T_\star<95{\rm eV}~\dot{E}_{34}^{-3/8}\nu_{\rm kHz}^{-1/4}.
\end{equation}
Here $\nu$ is the pulsar spin frequency. Now assume, crudely, that the surface temperature comes from the gamma-ray illumination and that those soft gamma-rays that are emitted near the light cylinder towards the star, are able to strike the stellar surfaces. 
We get
\begin{equation}
4\pi R_0^2\sigma T_\star^4\sim \frac{\pi R_0^2}{4\pi R^2}\epsilon \dot{E}.
\end{equation}
With $\epsilon \sim 0.2$, we obtain
\begin{equation}
T_\star\sim 36{\rm eV}~\dot{E}_{34}^{1/4}\nu_{\rm kHz}^{1/2},
\end{equation}
and Eq.(\ref{E:wpul}) gives the criterion for a strong rocket:
\begin{equation}
\dot{E}_{34}<4.7~\nu_{\rm kHz}^{-6/5}
\end{equation}
Here $\dot{E}_{34}\equiv \dot{E}/(10^{34}\hbox{erg/s})$.
From the spindown formula Eq.~(\ref{spitkovsky}), taking for concreteness $\theta=\pi/4$ and $R_0=12\hbox{km}$, we get
\begin{equation}
    \dot{E}_{34}=100 B_{10}^2 \nu_{\rm kHz}^3.
\end{equation}
From the above two equations, we get the necessary condition for the pulsar to be weak:
\begin{equation}
    \nu\lesssim \nu_{\rm weak}=480 B_{10}^{-10/21}\hbox{Hz}.
    \label{nucrit}
\end{equation}

\begin{figure}[htbp]
    \centering
    \includegraphics[width=\columnwidth]{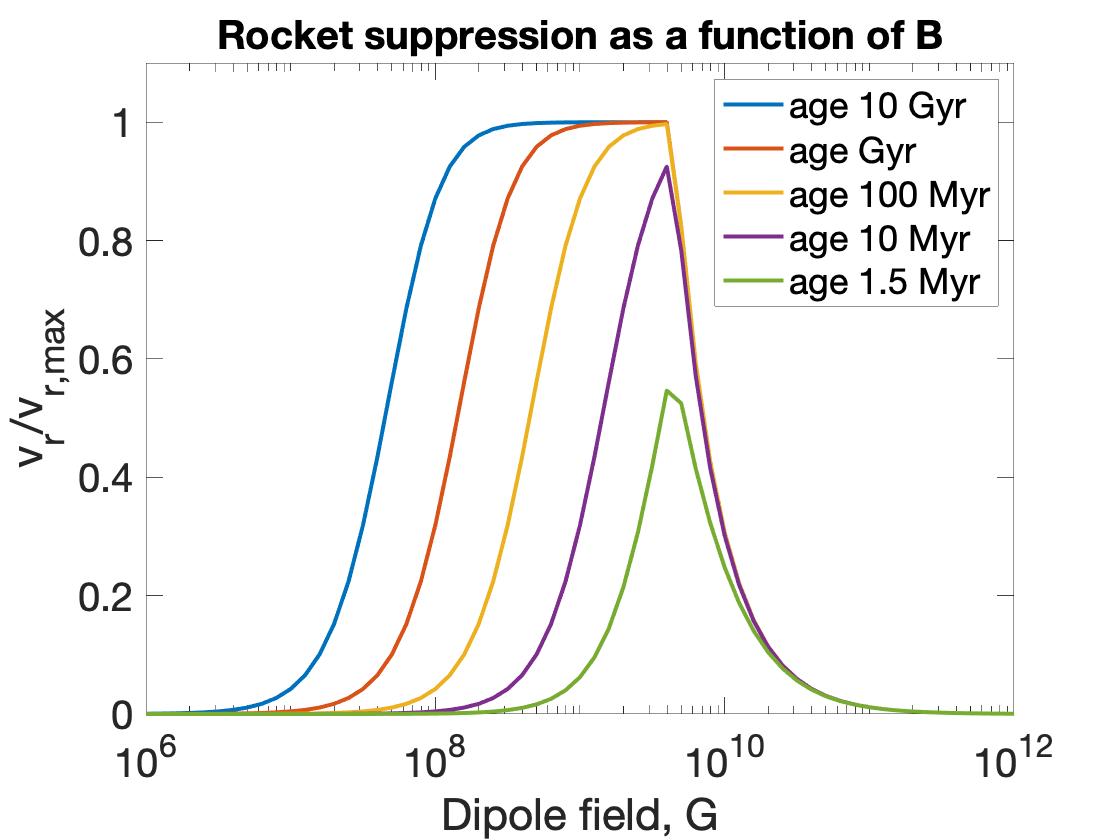}
    \caption{ The fraction of the maximal rocket velocity achievable for a pulsar as a function of its dipole magnetic field.}
    \label{fig:rocket_suppression}
\end{figure}

We now estimate the rocket velocity achievable for a pulsar with a given dipole field $B$. We shall crudely assume that the pulsar is force-free and the rocket force is zero when $\nu>\nu_{\rm weak}$. We (optimistically) initialize a pulsar at the spin frequency $\nu_0=\min(716~\hbox{Hz},\nu_{\rm weak})$ and assume that it spins down for time $T$, bringing it to the rotational frequency
\begin{equation}
\nu(T)=\left(\nu_0^{-2}+{12\pi^2\mu^2 T\over I c^3}\right)^{-1/2}.
\label{newfreq}
\end{equation}
For the purposes of our estimate we use $I=(2/5)M R_0^2$, $\mu=(1/2)B R_0^3$, and $M=1.5 M_{\odot}$, giving 
\begin{equation}
\nu(B,T)=\left[\nu_0^{-2}+(419\hbox{Hz})^{-2}B_{10}^2 T_{\rm Myr}\right]^{-1/2}.
\end{equation}
We now estimate what fraction $\eta$ of the maximum rocket velocity given by Eq.~(\ref{maximumrocket}) the pulsar can achieve:
\begin{equation}
\eta={\nu_0^3-\nu(B,T)^3\over (716\hbox{Hz})^3}.
\label{eta}
\end{equation}

The ratio $\eta$ is plotted in Figure 6. We see that to get a substantial rocket velocity, we need the dipole field $B\lesssim 10^{10}\hbox{G}$ and the pulsar age $T\gtrsim \hbox{Myr}$. These values are atypical for young pulsars, but are typical for old recycled millisecond pulsars. However,  recycling would be surprising in the context of Gaia binaries, and thus potentially we are dealing with a new class of neutron stars. 

'

\section{Discussion}
Since Gaia's discovery last year, we are dealing with a new population of neutron stars: invisible in electromagnetic radiation, likely very old, and located in wide binaries with a broad eccentricity distribution. We searched for evidence of the electromagnetic rocket force that would have acted on the neutron stars over their lifetimes, and found that for a large class of the initial configurations and natal kick distributions, the rockets of $v_r\gtrsim 30\hbox{km/s}$ are required to explain the binaries' eccentricities.  However, we also found some models of the binaries' initial conditions and natal kicks that produce the eccentricities and semimajor axes consistent with Gaia's observations. These models assume that 1. The binaries are initially wide and that they do not substantially shrink when the neutron star progenitors depart from the main sequence, and 2. The kick distribution allows substantial support at the low-velocity end. Perhaps new observations or a better understanding of the natal kicks would allow us to definitively determine whether such models are ruled out.

We have also dramatically revised electromagnetic rocket theory. We proved that the rocket force for a Force-Free magnetosphere is identically zero. This means that substantial kicks can only be developed by ``weak'' pulsars, which lack sufficient pair production at the light cylinder. Our crude estimates show that only old ($\gtrsim 1\hbox{Myr}$) pulsars with initially rapid spins ($\gtrsim 500\hbox{Hz}$), the small dipole field $B\lesssim 10^{10}$G and a relatively large quadrupole can achieve substantial rocket velocities. These are the characteristics common for millisecond pulsars, but these are not naturally expected 
to be found 
in wide binaries. Interestingly, there is evidence that some Central Compact Objects (CCO) in supernova remnants are weakly magnetic ($<10^{10} - 10^{11}$G) neutron stars with strongly non-dipolar fields [see \cite{2017JPhCS.932a2006D} and references therein]. One possible explanation for such weak fields is that the initially stronger fields were buried by the material accreted from  fall-back discs \citep{1989ApJ...346..847C, 1995ApJ...440L..77M, 1999A&A...345..847G}. The accretion could have also spun up the neutron stars; however, the highest measured rotational frequency in a CCO does not exceed $10$Hz \citep{2009ApJ...695L..35G}. It is therefore unclear whether there exists a natural astrophysical scenario for producing pulsar rockets in wide Gaia neutron star binary systems.

\section*{Acknowledgements}

We thank Ryosuke Hirai for introducing us to the problem and for insightful feedback on the draft of the paper, Kareem El-Badry for assistance with the Gaia selection function, Veome Kapil for providing the pulsar velocity data, and Andrei Beloborodov and Ashley Bransgrove for useful discussions. VB acknowledges support from the NASA Hubble Fellowship grant HST-HF2-51548.001-A awarded by the Space Telescope Science Institute, which is operated by the Association of Universities for Research in Astronomy, Inc., for NASA, under contract NAS5-26555. YL's work on this subject is supported by the Simons Collaboration on Extreme Electrodynamics of Compact Sources (SCEECS) and by Simons Investigator Grant 827103.

\bibliography{ref}

\appendix

\section{Comparison of Transverse Velocity Distributions with Observations}
\label{sec:transverse}

\begin{figure}[htbp]
    \centering
    \includegraphics[width=0.8\columnwidth]{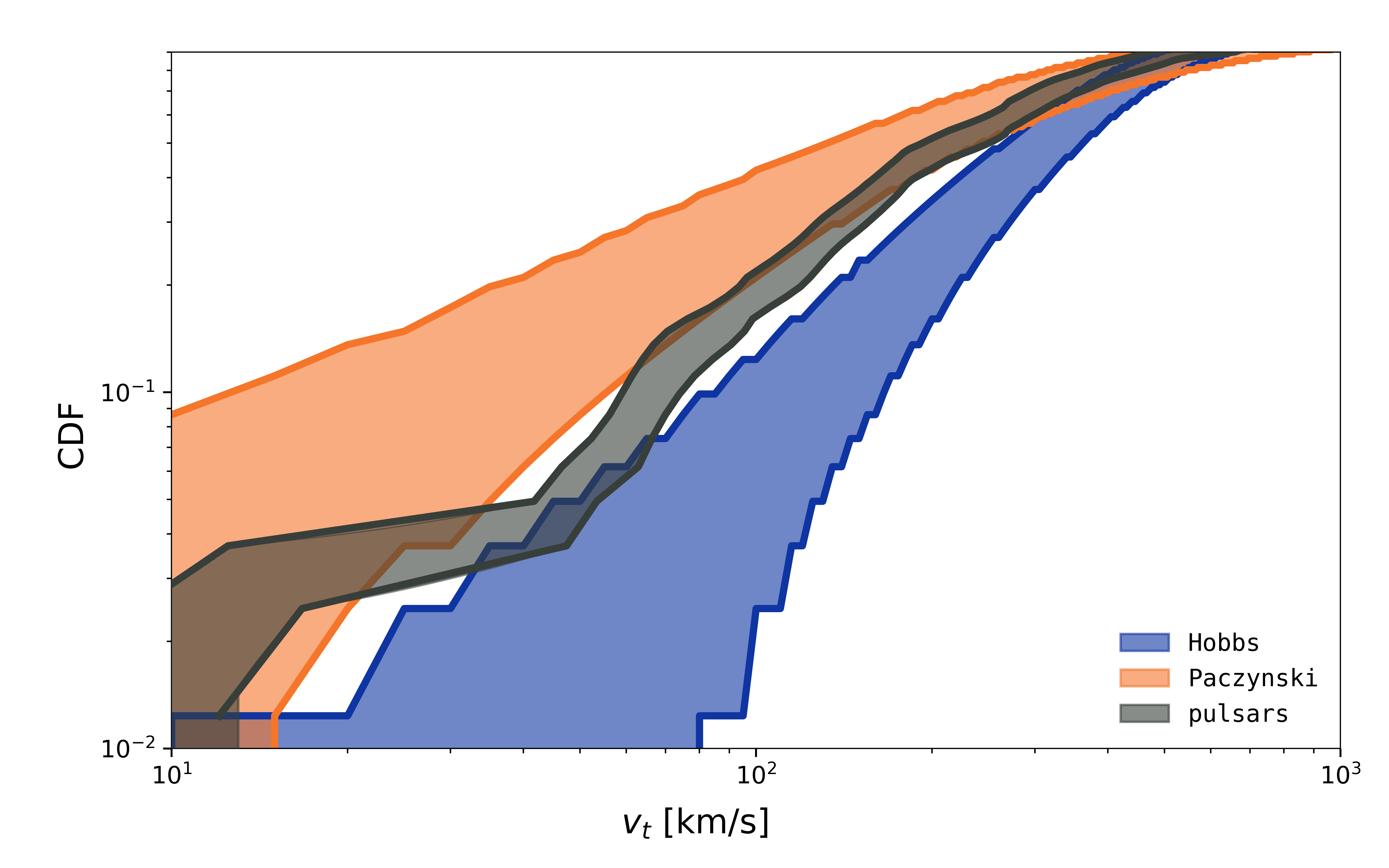}
    \caption{Comparison of the observed cumulative distribution of pulsar transverse velocities with predictions from two natal kick models. The solid black lines enclose the 95\% confidence interval derived from the pulsar data. The shaded blue and orange regions show the corresponding 95\% confidence intervals for predictions from the Hobbs and  Paczynski kick distributions, respectively.}
    \label{fig:transverse}
\end{figure}

Let us assume a population of pulsars, each with a 3D velocity vector $\vec{v}$. The magnitude of this vector, $v = |\vec{v}|$, follows the probability distribution $p(v)$. Our goal is to determine the probability distribution, $p(v_t)$, of the transverse velocity, which is the magnitude of the pulsar's velocity projected onto the plane of the sky. 

For any single pulsar with a 3D speed $v$, its velocity vector $\vec{v}$ forms an angle $\theta$ with our line of sight. The magnitude of the velocity projected onto the sky plane is $v_t = v \sin \theta$. For a fixed 3D speed $v$, the conditional probability distribution of its projection, $v_t$, is:
\begin{equation}
\label{eq:conditional_probability_vt}
p(v_t | v) = \frac{v_t}{v \sqrt{v^2 - v_t^2}} \quad \text{for } 0 < v_t < v
\end{equation}

The overall probability distribution of the transverse velocity, $p(v_t)$, is
\begin{equation}
\label{eq:overall_distribution_integral}
p(v_t) = \int_{v_t}^\infty p(v) \cdot p(v_t | v) \, dv = \int_{v_t}^\infty p(v) \frac{v_t}{v \sqrt{v^2 - v_t^2}} \, dv
\end{equation}

This fundamental relationship in Equation~\ref{eq:overall_distribution_integral} allows us to derive the probability distribution of the observed 2D transverse speeds, $p(v_t)$, from a given 3D speed distribution, $p(v)$. 

For the Hobbs Maxwellian model, the transverse velocity distribution is given by:
\begin{equation}
\label{eq:hobbs_transverse_distribution}
p(v_t) = \frac{v_t}{\sigma^2} e^{-\frac{v_t^2}{2\sigma^2}}
\end{equation}

For the Paczynski model, the resulting transverse velocity distribution is:
\begin{equation}
\label{eq:paczynski_transverse_distribution}
p(v_t) = \frac{1}{\sigma} \left( 2 - \frac{v_t (2v_t^2 + 3\sigma^2)}{(v_t^2 + \sigma^2)^{3/2}} \right)
\end{equation}

Next, we compare these theoretical models to observational data. We use the dataset of 81 pulsars from \citet{Willcox:2021kbg,Kapil:2022blf}, where transverse velocities were determined through Very Long Baseline Interferometry (VLBI). In Figure~\ref{fig:transverse}, we present the CDF of the transverse velocities for the 81 observed pulsars, including the 95\% confidence intervals derived from the data. We also plot the theoretical CDFs corresponding to the Hobbs (Eq.~(\ref{eq:hobbs_transverse_distribution})) and Paczynski (Eq.~(\ref{eq:paczynski_transverse_distribution})) transverse velocity distributions, along with their 95\% confidence intervals generated by drawing 81 samples from each respective distribution.

We find that the Hobbs model tends to underestimate the fraction of low-velocity systems, while the Paczynski model appears to overestimate them. The true distribution of the observed pulsar velocities seems to lie between these two models. Therefore, in this work, we will consider the Hobbs and Paczynski distributions as representing two plausible extremes for the underlying pulsar velocity distribution.

\section{Modeling Gaia Selection Effects}\label{sec:pdet}

Our goal is to model the detection probability $p_{\mathrm{det}}$ for astrometric binaries in Gaia DR3 as a function of orbital period $P_{\mathrm{orb}}$ and eccentricity $e$.  In this work, we use the tools and methodology developed by \citet{2024OJAp....7E.100E} to create a forward-model for
the Gaia DR3 astrometric binary catalog. They use the Gaia scanning law to simulate the observational cadence and scan angles at the level of individual one-dimensional astrometric measurements. This is coupled with a sequential fitting process that mirrors the cascade of models used in Gaia DR3: from single-star solutions, to accelerating models, to full Keplerian orbits.  
\begin{figure}[htbp]
    \centering
    \includegraphics[width=0.6\columnwidth]{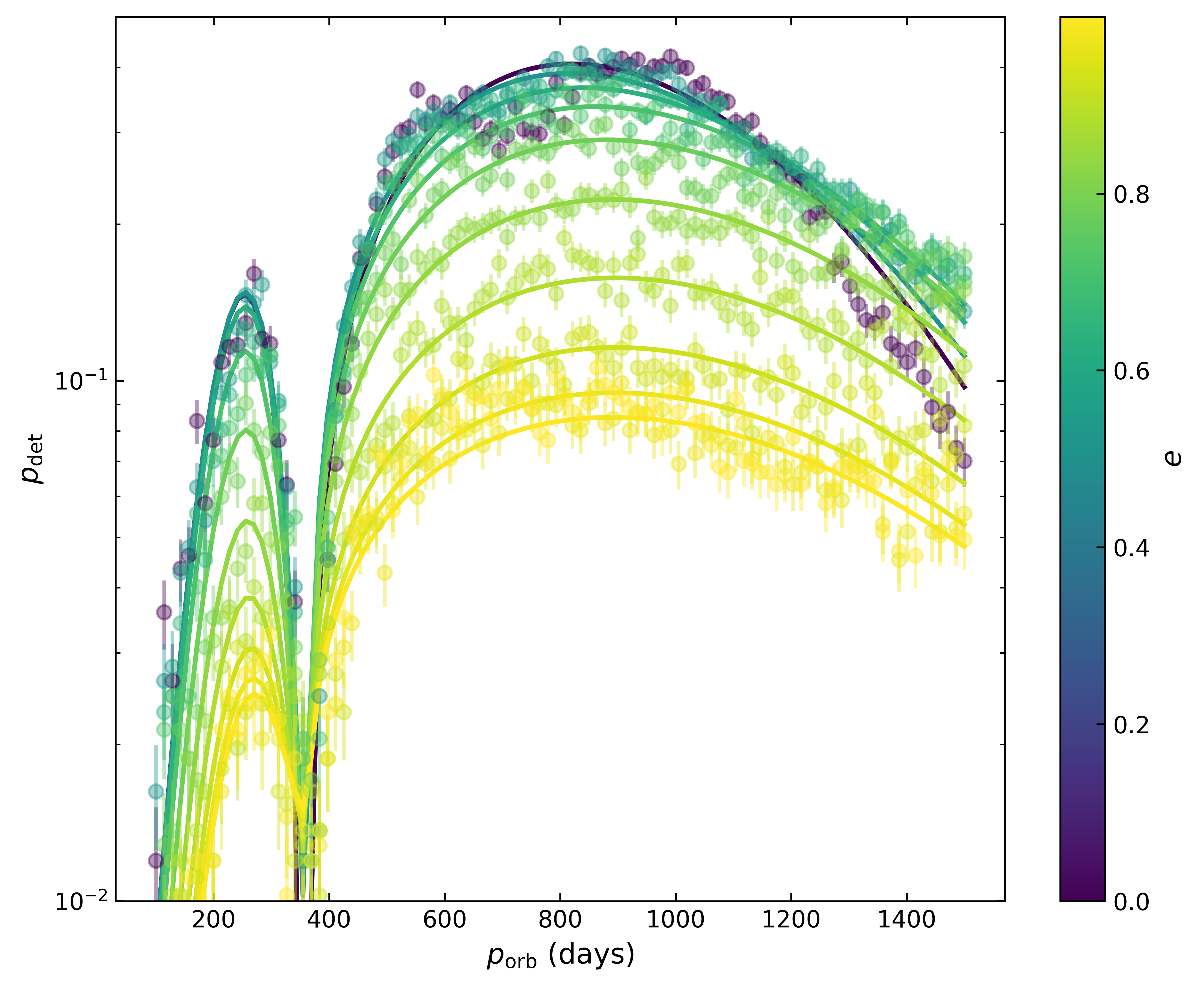}
    \caption{Comparison of the fitted detection probability $p_{\rm det}(P_{\rm orb}, e)$ (solid lines) with empirical estimates $\hat{p}_{\rm det}$ (points with error bars) computed from 1000 realizations per parameter set. Error bars represent binomial standard errors, $\Delta \hat{p}_{\rm det} = \sqrt{\hat{p}_{\rm det} (1 - \hat{p}_{\rm det}) / n}$. }
    \label{fig:p_det_1D}
\end{figure}

Our goal is to estimate the detectability of neutron star binaries in the Gaia DR3 astrometric catalog using the forward-modeling framework developed by \citet{2024OJAp....7E.100E} (2023). We simulate binary systems over a grid of orbital periods $ (P_{\mathrm{orb}}) $ and eccentricities $ (e) $. For each system, we sample the primary and companion masses, along with their absolute G-band magnitudes, based on neutron star binaries observed by Gaia. These binaries are then placed in the Milky Way using an exponential disk model, which assigns each system a sky position and distance. We also sample orbital parameters (such as including inclination, argument of periastron, and time of pericenter passage) randomly. At each point in the $ (P_{\mathrm{orb}}, e) $ grid, we generate $ n = 1000 $ mock realizations with varying orientations, sky positions, and distances.

We apply the Gaia DR3 astrometric detection criteria to each mock binary. The number of realizations that satisfy these criteria is denoted $ n_{\mathrm{det}} $, giving an empirical estimate of the detection probability:

\begin{equation}
\hat{p}_{\mathrm{det}} = \frac{n_{\mathrm{det}}}{n}
\end{equation}

To obtain a smooth and continuous model of $p_{\mathrm{det}}(P_{\mathrm{orb}, e}, e)$, we fit a parametric function to these estimates. The likelihood of observing $n_{\mathrm{det}}$ detections out of $n$ trials is modeled as a binomial distribution:

\begin{align}
\mathcal{L}(n_{\mathrm{det}} \mid p_{\mathrm{det}}(P_{\mathrm{orb}}, e)) 
&= \binom{n}{n_{\mathrm{det}}} \, p_{\mathrm{det}}(P_{\mathrm{orb}}, e)^{n_{\mathrm{det}}} \left(1 - p_{\mathrm{det}}(P_{\mathrm{orb}}, e)\right)^{n - n_{\mathrm{det}}}
\end{align}

To model the detection probability $p_{\mathrm{det}}(P_{\mathrm{orb}, e})$, we adopt a piecewise continuous function centered on $P_0 = 365$ days, with different behaviors below and above this point.
For $P_{\mathrm{orb}} < P_0$, the detection efficiency is modeled as a maxwellian function,

\begin{equation}
p_{\mathrm{det}}(P_{\mathrm{orb}}) = 
\left( A_1 \frac{(P_{\mathrm{orb}} - P_0)^2}{\sigma_1^3} + c \right) \exp\left(-\frac{(P_{\mathrm{orb}} - P_0)^2}{2\sigma_1^2} \right)
\end{equation}

Above this period $P_{\rm orb}>P_0$, the detection efficiency behaves like power-law modulated by a Gaussian envelope:
\begin{equation}
p_{\mathrm{det}}(P_{\mathrm{orb}}) = 
\left( A_2 \frac{(P_{\mathrm{orb}} - P_0)^\alpha}{\sigma_2^{\alpha + 1}} + c \right) \exp\left(-\frac{(P_{\mathrm{orb}} - P_0)^2}{2\sigma_2^2} \right)
\end{equation}

 The model is continuous at $P_0$, with separate parameters controlling the amplitude, width, and slope of the function on either side.  The parameters $A_1$, $\sigma_1$, $A_2$, $\sigma_2$, $\alpha$, and $c$ are functions of eccentricity $e$. We calibrated the model using Markov Chain Monte Carlo (MCMC) sampling to fit these parameters to empirical detection rates. The best-fit parameterizations are as follows:

 \begin{align}
A_1(e) &= 1.7059 + \frac{14.0569}{1 + \exp\left(16.9041 \cdot (e - 0.7742)\right)} \\
\sigma_1 &= 78.8048 \\
A_2(e) &= 73.3512 + \frac{257.3347}{1 + \exp\left(22.0893 \cdot (e - 0.8541)\right)} \\
\sigma_2(e) &= 741.9181 + \frac{-246.9119}{1 + \exp\left(7.4314 \cdot (e - 0.7547)\right)} \\
\alpha(e) &= 0.8391 - 0.2016 \cdot e \\
c(e) &= -0.0015 + 0.0162 \cdot e
\end{align}

In Figure~\ref{fig:p_det_1D}, we show the fitted detection probability model $ p_{\rm det}(P_{\rm orb}, e) $ alongside the empirical estimates $ \hat{p}_{\rm det} $, computed from $ n = 1000 $ realizations per grid point, with the associated uncertainties are estimated assuming binomial statistics $
\Delta \hat{p}_{\rm det} = \sqrt{(\hat{p}_{\rm det} (1 - \hat{p}_{\rm det}))/{n}}.$
The fitted model captures the overall structure of the detection probability surface well. However, we find that the simple parametric form used cannot fully reproduce the multiple local maxima observed at $ P_{\rm orb} > 365 $ days for circular binaries ($ e = 0 $), indicating that more complex models may be needed to capture the full richness of the detection landscape in this regime. As expected, binaries with orbital periods near one year exhibit large parallax uncertainties because their orbital motion becomes degenerate with parallactic motion. However, binaries with periods close to one year are more likely to be detected if they have eccentric orbits, since eccentricity helps break the degeneracy between parallax and orbital motion~\citep{2024OJAp....7E.100E}..

Overall, the detection probability decreases as eccentricity increases, with detection probability dropping off steeply for $e>0.6$ \citep{2024OJAp....7E.100E}.

\section{Bayesian Inference Methodology}
\label{sec:bayes}

We employ a Bayesian inference framework to constrain several physical parameters of our models using the observed Gaia data. This allows us to determine the posterior probability distribution of a given parameter, which we can denote generally as $\theta$. For example, $\theta$ might represent the minimum initial orbital separation, $a_{i, \text{min}}$ (as in Section~\ref{sec:bayesian_ai}), or the characteristic velocity of the rocket effect, $\sigma_r$ (as in Section~\ref{sec:bayesian_sigmar}).

We use forward-modeling to construct the likelihood function. For a range of possible values of the parameter $\theta$, we generate large synthetic populations of binary stars. We evolve each population through the supernova and rocket phases to determine the resulting distribution of final semi-major axes ($a_f$) and eccentricities ($e_f$). From these simulations, we numerically construct a smooth, two-dimensional probability density function, $p(a_f, e_f | \theta)$, which represents the probability of a surviving binary having specific orbital parameters for a given $\theta$. The total likelihood of our model for the entire observed dataset of $N=21$ Gaia binaries, $\{e_f^{(j)}, a_f^{(j)}\}$, is then the product of the individual probabilities:
\begin{equation}
\mathcal{L}(\{e_f^{(j)}, a_f^{(j)}\} | \theta) = \prod_{j=1}^N p(e_f^{(j)}, a_f^{(j)} | \theta)
\label{eq:likelihood}
\end{equation}

According to Bayes' theorem, the posterior probability of the parameter $\theta$ is proportional to this likelihood multiplied by a prior probability, $p(\theta)$:
\begin{equation}
p(\theta | \{e_f^{(j)}, a_f^{(j)}\}) \propto \mathcal{L}(\{e_f^{(j)}, a_f^{(j)}\} | \theta) \cdot p(\theta)
\label{eq:posterior}
\end{equation}

The resulting posterior distribution depends heavily on our choice of the prior, $p(\theta)$. A simple, uninformative prior would allow the likelihood to solely guide the inference. However, this can be misleading. The likelihood function, as defined above, evaluates how well a given $\theta$ reproduces the \textit{properties} of the observed binaries, but it does not account for the \textit{overall event rate}. A particular value of $\theta$ might produce a handful of systems that perfectly match the Gaia data, but if it simultaneously predicts that nearly all progenitor binaries are disrupted by the supernova or that the survivors are systematically undetectable by Gaia, it is not a physically plausible model.

To address this, we incorporate these critical selection effects directly into our prior. We construct a physically motivated prior that penalizes parameter values that yield an implausibly low number of detectable systems. We achieve this by defining the prior to be proportional to the fraction of systems that both survive the supernova and are detectable by Gaia, a quantity we denote as $f_{\text{surv, det}}(\theta)$:
\begin{equation}
p(\theta) \propto f_{\text{surv, det}}(\theta)
\end{equation}

This ensures that our inference favors models that not only match the observed data but also produce a reasonable yield of such systems.

For the specific case where our parameter is the minimum initial separation, $\theta = a_{i, \text{min}}$, we refine this prior slightly. It is standard practice to assume a log-uniform prior for a scale parameter like $a_{i, \text{min}}$ when there is no other information. To incorporate this, we define our prior as:
\begin{equation}
p(a_{i, \text{min}}) \propto \frac{f_{\text{surv, det}}(a_{i, \text{min}})}{a_{i, \text{min}}}
\end{equation}
In absence of any survival or detection biases ($f_{\text{surv, det}} = \text{constant}$), this prior correctly reduces to a log-uniform distribution.

\begin{figure}[htbp]
    \centering
    \includegraphics[width=\textwidth]{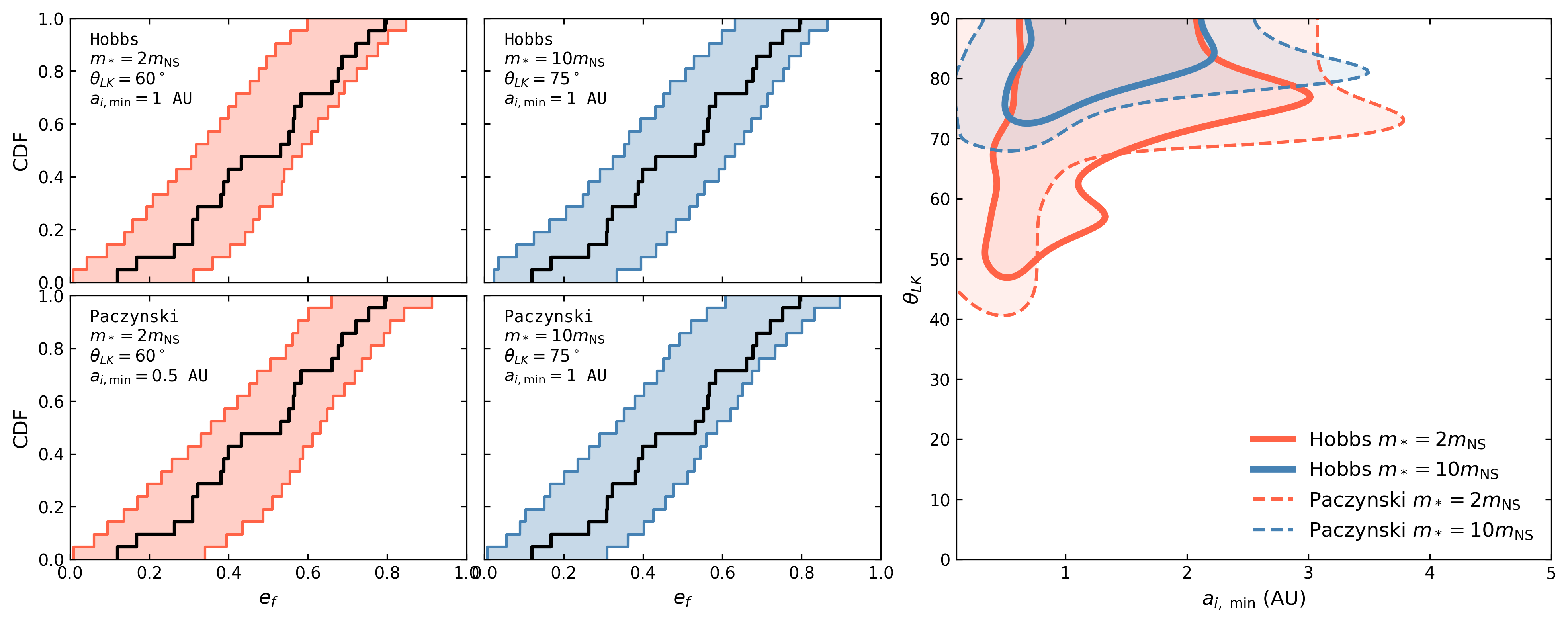}
    \caption{
        Left: CDFs of final orbital eccentricities ($e_f$) assuming misaligned kicks comparing synthetic neutron star binary populations with Gaia DR3 observations. All models assume initially circular orbits ($e_i = 0$) with initial semi-major axes $a_i$ log-uniformly distributed from the specified minimum to $100\,$AU. Each subplot shows the 99\% confidence interval (shaded regions) for different kick distributions and stellar progenitor masses. Top left: Hobbs kick distribution with low-mass progenitor ($m_* = 2m_{\rm NS}$, $\theta_{LK} = 60^\circ$, $a_{i,\rm min} = 1\,$AU). Top right: Hobbs kicks with high-mass progenitor ($m_* = 10m_{\rm NS}$, $\theta_{LK} = 75^\circ$, $a_{i,\rm min} = 1\,$AU). Bottom left: Paczynski kick distribution with low-mass progenitor ($m_* = 2m_{\rm NS}$, $\theta_{LK} = 60^\circ$, $a_{i,\rm min} = 0.5\,$AU). Bottom right: Paczynski kicks with high-mass progenitor ($m_* = 10m_{\rm NS}$, $\theta_{LK} = 75^\circ$, $a_{i,\rm min} = 1\,$AU). The observed CDF from Gaia DR3 neutron star binaries (black curves) is overlaid on each subplot, showing showing strong agreement with the all the models. The angle $\theta_{LK}$ represents the maximum opening angle between the natal kick velocity vector and the orbital angular momentum vector of the pre-supernova binary, meaning $\theta_{LK}=0^\circ$ indicates a kick perpendicular to the orbital plane, while $\theta_{LK}=90^\circ$ signifies an essentially isotropic kick.
        Right Panel: Regions of $a_{i,\rm min}$-$\theta_{LK}$ parameter space where synthetic populations are consistent with the Gaia DR3 eccentricity distribution. Solid lines represent contours for Hobbs kick distributions, while dashed lines show Paczynski kick distributions. Red contours correspond to low-mass progenitors ($m_* = 2m_{\rm NS}$) and blue contours to high-mass progenitors ($m_* = 10m_{\rm NS}$). All populations assume initially circular orbits.
    }
    \label{fig:non_polar_kicks}
\end{figure}

\section{Exploring Misalignment between Spin and Natal Kick}
\label{sec:opening_angle}

Thus far, our analysis has focused on two primary scenarios for natal kick directions: fully isotropic kicks or strictly polar kicks (aligned with the neutron star's spin , $\theta_{LK}=0^\circ$, assuming the spin  is aligned with the pre-supernova orbital angular momentum). We've seen that for initially circular orbits and a progenitor mass of $m_* = 2m_{\rm NS}$, purely polar kicks struggle to reproduce the observed Gaia eccentricity distribution without invoking a subsequent rocket phase. But what happens if the natal kick is neither perfectly isotropic nor perfectly polar, but instead occurs at some intermediate angle to the orbital angular momentum?

To study this, we consider scenarios in which the natal kick direction is isotropically distributed within a cone of half-opening angle $\theta_{LK}$, centered on the pre-supernova orbital angular momentum vector. For instance, let's study the case where $\theta_{LK} = 60^\circ$. We simulate synthetic populations under this assumption, using the Hobbs kick distribution, assuming initially circular orbits, wide initial separations ($a_i$ drawn from a log-uniform distribution in $[1, 10]$ AU), and a progenitor mass of $m_* = 2m_{\rm NS}$.

In Fig.~\ref{fig:non_polar_kicks} shows the resulting cumulative eccentricity distribution. The plot compares the synthetic population (with 99\% confidence intervals shaded) against the observed Gaia CDF (overplotted in black). Remarkably, under these conditions, we find that the observed Gaia eccentricity distribution is largely consistent with the model predictions, even without invoking rocket effects. Note that consideringt purely polar kicks ($\theta_{LK} = 0^\circ$) under similar initial conditions (circular, wide, $m_* = 2m_{\rm NS}$) were unable to adequately explain the observed eccentricities (as seen in Figure~\ref{fig:cdf_kick}, top row, red curves). This suggests that the kicks need to be misaligned from the purely polar direction to explain Gaia eccentrcities. In Fig.~\ref{fig:non_polar_kicks}, we also present additional scenarios using the Hobbs kick distribution and the Paczyński distribution for massive progenitor masses of $m_* = 2m_{\rm NS}$ and $m_* = 10m_{\rm NS}$. These cases also produce results that are consistent with the Gaia eccentricity distribution.

This shows that a minimum cone half-opening angle is required to reproduce the Gaia eccentricity distribution for a given separation distribution. For a set of initial orbital separations $a_i \in [a_{i, \rm min}, 100 {\rm AU}]$, we determine the minimum cone half-opening angle $\theta_{\rm LK}$ necessary to achieve consistency with the Gaia observations assuming initially circular orbits. This relationship is shown in the right column of Fig.~\ref{fig:non_polar_kicks}. Note that for Paczyński kicks, $a_{i, \rm min} \sim \mathcal{O}(10^{-2}\,{\rm AU})$, are consistent with the Gaia observations. However, it is important to emphasize that the majority of Gaia-detectable systems originate from larger values of $a_i$, even when the minimum allowed separation $a_{i, \rm min}$ is very small.

For relatively low-mass progenitors with $m_* = 2m_{\rm NS}$, eccentricity distributions consistent with Gaia witin 99\%  can be produced for minimium initial separations up to approximately $3$--$4\,\text{AU}$. In these cases, a minimum cone half-opening angle $\theta_{\rm LK}$ of around $\sim45^\circ$ is required. In contrast, for massive progenitors with $m_* = 10m_{\rm NS}$, a significantly larger minimum angle of at least $70^\circ$ is necessary to match the Gaia data. This effectively demands an almost isotropic kick distribution.

These  results clearly demonstrate that purely polar kicks or kicks with only small misalignments, as suggested by some radio observations, are incompatible with Gaia’s eccentricity measurements if binaries circularize prior to supernova.

\section{Inferring Rocket Velocities for Individual Gaia Binaries}
\label{sec:individual_rockets}

In Sec.~\ref{sec:gaia_rockets}, we studied the influence of rocket acceleration on population-level eccentricity distributions by varying the population-wide parameter $\sigma_r$ of the rocket velocity distribution. While this allowed us to understand the general role of rockets in shaping binary orbital properties, it does not address the question of whether individual Gaia-detected neutron star binaries require a rocket to explain their orbital configurations and if so, how strong.

\begin{figure}[h!]
\centering
\includegraphics[width=0.6\textwidth]{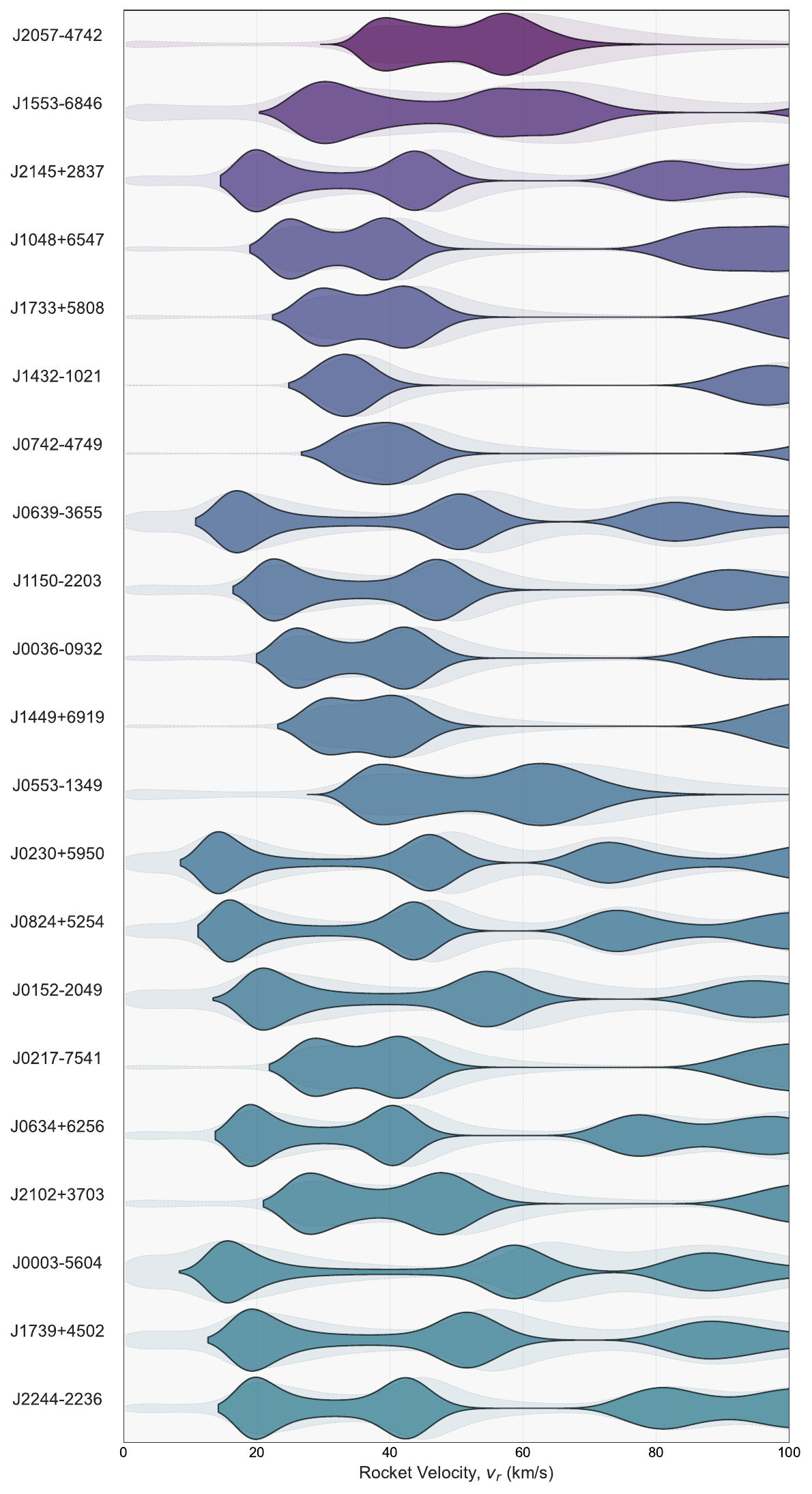}
\caption{
Posterior probability distributions of rocket velocity ($v_r$) for individual Gaia neutron star–stellar binaries. Each panel corresponds to one binary. Solid violin plots assume compact pre-supernova orbits (reduced orbit), while dashed, semi-transparent violins represent wide orbits. The distributions were inferred via Bayesian framework under a Hobbs natal kick distribution, thermal eccentricity distribution, and a progenitor mass of $m_* = 2m_{\rm NS}$. Color indicates orbital period: darker colors denote shorter-period systems, while brighter colors correspond to longer-period systems. Systems with low eccentricity and reduced pre-SN orbits generally require a minimum nonzero $v_r$ to match observations, while wide orbit configurations allow for a broader range of rocket velocities, often including $v_r = 0$.
}
\label{fig:rocket_violin}
\end{figure}

In this section, we instead infer rocket velocities on a per-source basis. That is, we fix the natal kick distribution (e.g., Hobbs), progenitor mass ($m_* = 2 m_{\rm NS}$), and pre-supernova orbital distributions (either wide or reduced), and ask: What rocket velocity $v_{r,j}$ would best explain the observed orbital eccentricity $e_{f,j}$ of each individual binary $j$?

For each Gaia binary, we calculate the likelihood of observing the system’s current orbital parameters (eccentricity and period) given a rocket velocity $v_r$. Rather than marginalizing over a distribution of $v_r$ values parameterized by $\sigma_r$, we treat $v_r$ as an explicit free parameter per system and adopt a Bayesian framework to infer the most likely rocket velocity for each observed system. 
Figure~\ref{fig:rocket_violin} shows the resulting posterior distributions of rocket velocity $v_r$ for each Gaia neutron star binary, plotted as violin plots. Each panel corresponds to a different binary. The solid violins represent the case where the binary underwent significant orbital shrinkage prior to supernova ("reduced orbit"), while the dashed, semi-transparent violins correspond to the wide orbit case, where no CE-induced shrinkage is assumed. All results assume a Hobbs natal kick distribution and thermal pre-SN eccentricity. Color encodes the orbital period of the binary: darker shades indicate shorter periods and lighter shades correspond to longer periods.

\textit{Reduced Orbit:} The inferred $v_r$ distributions are generally more compact and peaked, with some binaries requiring a sharply defined nonzero rocket velocity to reconcile the high predicted post-SN eccentricities with their moderate observed values. This tightness reflects a form of dynamical fine-tuning: in reduced orbits, natal kicks alone would yield very high eccentricities ($e_f \gtrsim 0.95$), so the rocket must act to precisely lower $e_f$ into the observed range. Additionally, for many systems there exists a minimum required $v_r$—the posterior has no support near $v_r = 0$.

\textit{Wide Orbit:} Here, the inferred rocket velocities are typically lower, and the posterior distributions often show significant support down to $v_r = 0$, especially for binaries with higher observed eccentricities. This reflects the fact that in wide orbits, natal kicks alone are often sufficient to produce moderate post-SN eccentricities. In several cases such as those with large eccentricities, a rocket is not required, though systems with small eccentricities prefer some rockets.

Overall, these per-source inferences provide a consistency check on the rocket hypothesis. They show that while not all Gaia binaries require a rocket, some fraction, especially those with lower eccentricities and reduced pre-SN orbits might favor a nonzero $v_r$.

\end{document}